\documentclass[pre,showpacs,showkeys,preprintnumbers,amsmath,amssymb,superscriptaddress,twocolumn]{revtex4}
\usepackage{graphicx}
\usepackage{amsfonts}
\usepackage{url}
\usepackage{epstopdf}
\usepackage{color}
\usepackage{tikz-cd}
\usepackage[colorlinks=true,linkcolor=blue,citecolor=red]{hyperref}%

\def\L{\mathcal L}

\def\ve{\varepsilon}

\def\pa{{\partial\Omega}}

\def\R{{\mathbb R}}

\def\C{{\mathbb C}}

\def\ctan{\mathrm{ctan}}

\def\kkappa{\bar{\kappa}}
\def\llambda{\bar{\lambda}}
\def\mmu{\bar{\mu}}

\def\kon{k_{\rm on}}

\def\nmax{n_{\rm max}}

\begin{document}

\title{Reversible reactions controlled by surface diffusion on a sphere}

\author{Denis~S.~Grebenkov}
 \email{denis.grebenkov@polytechnique.edu}
\affiliation{
Laboratoire de Physique de la Mati\`{e}re Condens\'{e}e (UMR 7643), \\ 
CNRS -- Ecole Polytechnique, IP Paris, 91128 Palaiseau, France}

\date{\today}

\begin{abstract}
We study diffusion of particles on the surface of a sphere toward a
partially reactive circular target with partly reversible binding
kinetics.  We solve the coupled diffusion-reaction equations and
obtain the exact expressions for the time-dependent concentration of
particles and the total diffusive flux.  Explicit asymptotic formulas
are derived in the small target limit.  This study reveals the strong
effects of reversible binding kinetics onto diffusion-mediated
reactions that may be relevant for many biochemical reactions on cell
membranes.
\end{abstract}

\pacs{02.50.-r, 05.40.-a, 02.70.Rr, 05.10.Gg}



\keywords{surface diffusion, reaction rate, cellular membrane, first passage time, reversible reaction}

\maketitle

\section{Introduction}

Various vital processes occur on the cell membrane, examples ranging
from cell signaling to ionic and molecular transfer \cite{Albert}.  In
some cases, proteins diffuse on the cell membrane and search for
specific receptors to trigger signal transduction via a cascade of
biochemical reactions.  For instance, G-protein-coupled receptors
transmit their signals to the cell interior by interacting with G
proteins and thus mediate the biological effects of many hormones and
neurotransmitters \cite{Sungkaworn17}.  A comprehensive theory of
these crucial phenomena requires a better understanding of diffusive
processes on two-dimensional surfaces and, in particular, on a
spherical surface.  In this direction, Bloomfield and Prager studied
the impact of rotational diffusion on tail fiber attachment on
bacteriophage T4 \cite{Bloomfield79}.  Assuming that the fiber tip
diffuses freely over the surface of the sphere until it reaches a
baseplate site, they computed the first two moments of the
distribution of reaction times.  Chao {\it et al.}  investigated
surface diffusion into a trap as a promising passive mechanism of
localization of cell membrane components \cite{Chao81} (see also
\cite{Weaver83}).  They derived the concentration of components by
solving the time-dependent diffusion equation in the presence of a
perfectly absorbing sink.  Linderman and Lauffenburger analyzed
trapping of diffusing receptors into tubules as a mechanism of
intracellular receptor/ligand sorting \cite{Linderman86}.  Sano and
Tachiya treated a more general case of imperfect sinks \cite{Sano81},
for which they obtained the mean reaction time and discussed an
exponential approximation for the survival probability.  Berg
considered the effect of weak binding of a diffusing particle to the
nonspecific surface around a target on the sphere \cite{Berg85}.  The
resulting surface-mediated diffusion, in which three-dimensional
diffusion in the bulk is coupled to two-dimensional diffusion on the
surface, facilitates the target search.  The optimality of such
intermittent processes was further investigated
\cite{Benichou10,Benichou11,Rojo11,Rupprecht12a,Rupprecht12b}.
Berg and Purcell estimated the reaction rate controlled by bulk
diffusion toward multiple small patches evenly distributed over the
surface of a sphere \cite{Berg77}.  This seminal paper and the
associated homogenization concept were further extended by many
authors
\cite{Zwanzig90,Zwanzig91,Berezhkovskii04,Berezhkovskii06,Muratov08,Dagdug16,Lindsay17,Bernoff18a,Bernoff18b}.
Shoup {\it et al.} devised an efficient approximation to compute the
reaction rate from the bulk to an active circular region on the
surface of a sphere, even in the presence of rotational motion
\cite{Shoup81}.  This approximation was further extended to compute
the mean first-passage time (MFPT) from the bulk \cite{Grebenkov17}.
Singer {\it et al.}  generalized the small-target asymptotic behavior
of the MFPT to two-dimensional manifolds \cite{Singer06c}.  In turn,
Coombs {\it et al.}  derived the asymptotic behavior of the MFPT for a
diffusing particle confined to the surface of a sphere, in the
presence of many partially absorbing traps of small radii
\cite{Coombs09}.  Pr\"ustel and Tachiya considered the problem of
reversible binding reactions on several two-dimensional domains,
including the spherical case \cite{Prustel13}.  By employing
convolution relations between the survival probabilities for
reversible and irreversible reactions, they derived an exponential
approximation for these probabilities.

In this paper, we further extend these former works by considering
surface diffusion of particles on a sphere toward a partially reactive
circular target with partly reversible binding kinetics
(Sec. \ref{sec:model}).  By generalizing the approach by Chao {\it et
al.} \cite{Chao81}, we provide in Sec. \ref{sec:solution} the exact
solution for the concentration of diffusing particles in both Laplace
domain and time domain.  In Sec. \ref{sec:discussion}, we illustrate
the properties of our solution, reveal the strong effects of
reversible binding kinetics, and discuss implications for biochemical
reactions on cell membranes.  Explicit asymptotic formulas in the
small target limit are also presented.  Section \ref{sec:conclusion}
concludes the paper with the summary of results and perspectives for
future research.  Technical derivations are reported in Appendices.

\section{Model}
\label{sec:model}

\begin{figure}
\begin{center}
\includegraphics[width=40mm]{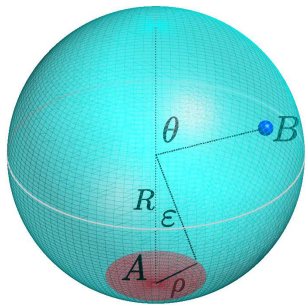} 
\end{center}
\caption{
Diffusion of a particle $B$ (blue point) on the surface of a sphere
toward a particle $A$ (a target of angular size $\ve$ or,
equivalently, of radius $\rho = R \sin \ve$) fixed at the South pole
(dark red region).}
\label{fig:sphere}
\end{figure}

We consider the following model: a single particle $A$ of radius
$\rho$ is fixed at the South pole of the sphere of radius $R$ whereas
point-like particles $B$ diffuse on the surface of that sphere with
the diffusion coefficient $D$ (Fig. \ref{fig:sphere}).  In what
follows, we call the particle $A$ a target (or a sink, or a trap).
This model can also describe the situation when the particle $A$
diffuses with the coefficient coefficient $D_A$ or/and when the
particles $B$ have a finite radius $\rho_B$.  In this case, $D$ is
replaced by $D + D_A$ while $\rho$ is replaced by $\rho + \rho_B$.
However, the particles $B$ are still assumed to be independent,
without any mutual interaction (even without the excluded volume
constraint).  In practice, it means that the concentration of
particles $B$ should be rather dilute.

We study a common diffusion-influenced reaction
\begin{equation}
\label{eq:kinetics}
\begin{tikzcd}[every arrow/.append style={shift left}]
 A + B \arrow{r}{\kon} & AB \arrow{l}{{\lambda}} \overset{\mu}{\longrightarrow} A + C .
  \end{tikzcd}
\end{equation}
Once a particle $B$ comes in contact with the target particle $A$, it
can either bind with the target, or continue its diffusive motion on
the surface.  The probability of the binding event is determined by
the reactivity $\kappa$ (in units m/s) in a standard way
\cite{Filoche99,Grebenkov03,Grebenkov06,Grebenkov07a,Singer08}.  The
latter can also be related to the conventional association rate
constant $\kon$ (in units m$^2$/s/mol) as
\begin{equation}
k_{\rm on} = \kappa N_A (2\pi \rho) ,
\end{equation}
where $N_A \simeq 6\cdot 10^{23}$~1/mol is the Avogadro number, and
$2\pi \rho$ is the target perimeter \cite{Shoup82,Lauffenburger}
(another notation, $k_d = k_{\rm on}/N_A = 2\pi \rho \kappa$, was
employed in \cite{Agmon90,Prustel13}).  A finite reactivity $\kappa$
mimics conformational or energy activation barrier that the particles
$A$ and $B$ have to overcome to form a metastable complex $AB$.  Two
rates $\lambda$ and $\mu$ allow us to distinguish three scenarios of
irreversible ($\lambda = 0$), reversible ($\lambda >0$, $\mu = 0$),
and partly reversible ($\lambda > 0$, $\mu > 0$) binding.  When the
dissociation (or desorption) rate $\lambda$ is zero, the particle $B$
never resumes its diffusion, being either chemically transformed into
another particle $C$ with the rate $\mu$, or staying bound forever to
the target (if $\mu = 0$).  In the former case, the target $A$ plays
the role of a catalyst for the transformation of $B$ into $C$.  When
$\lambda > 0$ and $\mu = 0$, the formation of the complex $AB$ is
fully reversible, i.e., the particle $B$ remains bound for a random
exponentially distributed time, determined by the dissociation rate
$\lambda$, and then resumes its surface diffusion until the next
binding.  This mechanism accounts for reversible reactions 
\cite{Tachiya80,Agmon84,Agmon90,Kim99}.  In turn, the last case
$\lambda > 0$ and $\mu > 0$ describes partly reversible scenario when
the metastable complex $AB$ can dissociate either into $A + B$ with
the rate $\lambda$ (in which case $B$ resumes its diffusion until the
next binding), or into $A + C$ with the rate $\mu$.  Note that the
rate $\lambda$ was also denoted as $k_{\rm off}$ or $k_a$.

In spherical angular coordinates $(\theta,\varphi)$, the diffusion
domain on the surface of the sphere is $\Omega = \{ (\theta,\varphi)
~:~ 0 \leq \theta < \pi - \ve, ~0 \leq \varphi < 2\pi \}$, whereas the
target surface is $\pa = \{ (\theta,\varphi) ~:~ \theta = \pi - \ve,
~0 \leq \varphi < 2\pi \}$, where $\ve$ is the angular size of the
target: $\rho = R \sin \ve$.  We introduce the surface concentration
of particles $B$ in a point $(\theta,\varphi)$, $c(\theta,\varphi,t)$
(in mol/m$^2$) that obeys the diffusion equation
\begin{equation}  \label{eq:v_diff}
\partial_t c = D \Delta c  \quad \textrm{in}~\Omega,
\end{equation}
where $\Delta$ is the Laplace-Beltrami operator on the sphere:
\begin{equation}
\Delta = \frac{1}{R^2} \biggl(\frac{1}{\sin\theta} \partial_\theta \sin \theta \partial_\theta 
+ \frac{1}{\sin^2\theta} \partial^2_\varphi\biggr).
\end{equation}
Even though a Dirac point-like source would yield a more general
description, we restrict the analysis to the uniform initial
concentration of particles $B$:
\begin{equation}
c(\theta,\varphi,t=0) = C_0. 
\end{equation}
This assumption, which is rather common for applications, simplifies
technical computations.  In fact, the axial symmetry of the domain and
of the initial condition ensures that the solution
$c(\theta,\varphi,t)$ does not depend on the azimuthal angle $\varphi$
that we drop in the following.  It is worth noting that the developed
method can be extended to a general setting without axial symmetry.

The diffusion equation (\ref{eq:v_diff}) is completed by the Robin
boundary condition on the target boundary $\pa$ 
\begin{equation}  \label{eq:v_BC}
- D (\partial_n c)_{|\pa} = \kappa \, c_{|\pa} - \lambda \frac{n_T}{2\pi \rho} \,,
\end{equation}
where $\partial_n = (1/R) \partial_\theta$ is the normal derivative
directed toward the target, and $n_T(t)$ (in mol) is the amount of
particles $B$ bound to the target at time $t$.  Here, $n_T/(2\pi
\rho)$ can be understood as the density of bound particles along the
target boundary of length $2\pi \rho$.  In chemical physics, the
relation (\ref{eq:v_BC}) is called ``back-reaction'', ``generalized
radiation'', or ``generalized Collins-Kimball'' boundary condition
\cite{Goodrich54,Agmon84,Kim99,Prustel13}.  It states that the net
diffusive flux density of particles $B$ toward the target (the
left-hand side) is equal to the reactive flux density, $\kappa
c_{|\pa}$, minus the flux density related to the dissociation of the
metastable complex $AB$ (the last term).  The reactivity $\kappa$
characterizes the difficulty for a particle $B$ to overcome an energy
activation barrier for binding to the target
\cite{Collins49,Sano79,Sapoval94} (see also a recent overview in
\cite{Grebenkov19c,Grebenkov19d}).  The limit $\kappa = \infty$
describes an immediate reaction upon the first encounter (no
activation barrier), whereas $\kappa = 0$ corresponds to no reaction
(infinite activation barrier) on reflecting boundary.  The
concentration is also required to be regular (not diverging) at the
North pole that will serve as the second boundary condition at $\theta
= 0$, see below.

In turn, the amount of particles $B$ bound to the target, $n_T(t)$,
obeys the first-order differential equation:
\begin{equation} \label{eq:vT_eq}
\partial_t n_T = J(t) - \mu \, n_T,
\end{equation}
with the initial condition $n_T(t=0) = 0$, where
\begin{equation} \label{eq:Jt_def}
J(t) = \int\limits_\pa dl \, \bigl(-D (\partial_n c)_{|\pa}\bigr) = -2\pi \rho D (\partial_n c)_{|\pa}
\end{equation}
is the total diffusive flux onto the target, and the second term
accounts for irreversible transformation of the metastable complex
$AB$ into $A$ and $C$.  Once $J(t)$ is found, one gets immediately
\begin{equation}  \label{eq:nT}
n_T(t) = \int\limits_0^t dt' \, e^{-\mu (t-t')} \, J(t').
\end{equation}
In the case $\mu > 0$, the amount $n_C(t)$ of produced particles $C$
obeys an ordinary differential equation,
\begin{equation}  \label{eq:nC}
\partial_t n_C(t) = \mu \, n_T(t)  ,
\end{equation}
with the initial condition $n_C(0) = 0$.  Once $n_T(t)$ is found,
$n_C(t)$ follows immediately by integration.

\section{Solution}
\label{sec:solution}

In this section, we solve the coupled partial differential equations
introduced in Sec. \ref{sec:model}.  As it is common for diffusion
problems, we first obtain the solution in Laplace domain (for
Laplace-transformed quantities) and then proceed to the solution in
time domain via Laplace transform inversion.

\subsection{Solution in Laplace domain}

Setting
\begin{equation}
x = \cos\theta, 
\end{equation}
the Laplace-Beltrami operator can be written as
\begin{equation}  \label{eq:Ldef}
\Delta = \L/R^2, \qquad \L = \partial_x (1-x^2)\partial_x.  
\end{equation}
In order to get rid of the derivative with respect to time $t$, the
Laplace transform is performed,
\begin{subequations}
\begin{eqnarray}
\tilde c(x,s) &=& \int\limits_0^\infty dt \, e^{-ts} \, c(x,t) , \\
\tilde n_T(s) &=& \int\limits_0^\infty dt \, e^{-ts} \, n_T(t)
\end{eqnarray}
\end{subequations}
(throughout the paper, tilde denotes Laplace-transformed quantities).
The new functions $\tilde c(x,s)$ and $\tilde n_T(s)$ satisfy
\begin{subequations}
\begin{eqnarray} \label{eq:auxil1}
s \, \tilde c(x,s) - \frac{D}{R^2} \, \L \,\tilde c(x,s) &=& C_0 , \\  
\label{eq:auxil2}
(s + \lambda + \mu) \, \frac{\tilde n_T(s)}{2\pi \rho} &=& \kappa \, \tilde c(a,s) , \\  \label{eq:auxil3}
\hspace*{-5mm}
\frac{D}{R} \sqrt{1-a^2} \bigl(\partial_x \tilde c(x,s)\bigr)_{x=a} = \kappa \, \tilde c(a,s) &-& \frac{\lambda \tilde n_T(s)}{2\pi \rho} \,,  
\end{eqnarray}
\end{subequations}
where we used $\partial_n = - (1/R) \sin\theta\, \partial_x$, the initial
condition $n_T(t=0) = 0$, and
\begin{equation}
a = \cos (\pi-\ve) = - \cos\ve = - \sqrt{1 - (\rho/R)^2}
\end{equation}
is the location of the target surface $\pa$.  As Eq. (\ref{eq:auxil2})
relates $\tilde n_T(s)$ and $\tilde c(a,s)$, one can rewrite
Eq. (\ref{eq:auxil3}) as Robin boundary condition
\begin{equation}  \label{eq:auxil1a}
\bigl(\partial_x \tilde c(x,s)\bigr)_{x=a} = q_s \, \tilde c(a,s) , 
\end{equation}
with
\begin{equation}  \label{eq:qs}
q_s = q \, \frac{s+\mu}{s+\lambda + \mu} \,,
\end{equation}
where
\begin{equation}  \label{eq:q}
q = \frac{\kkappa}{\sqrt{1-a^2}} \geq 0 \,, \qquad \kkappa = \frac{\kappa R}{D} 
\end{equation}
(throughout the paper, bar denotes dimensionless quantities rescaled
by $R$ and $D$).  In other words, the above coupled differential
equations for $c$ and $n_T$ are decoupled in Laplace domain and
reduced to a single differential equation
\begin{equation}  \label{eq:eigen_prob}
\bigl(R^2 s/D - \L\bigr) \tilde c = \frac{R^2 C_0}{D} \qquad (a < x < 1) ,
\end{equation}
subject to the Robin boundary condition (\ref{eq:auxil1a}) at $x = a$
(the target), and the regularity condition at $x = 1$ (the North
pole).  This implies that the reversible kinetics can be described in
Laplace domain as irreversible one, with a ``frequency''-dependent
reactivity
\begin{equation}
\kappa(s) = \kappa \, \frac{s+\mu}{s+\lambda+\mu} \,.  
\end{equation}
In time domain, this would result in a convolution-type Robin boundary
condition with an ``effective'' time-dependent reactivity.  We note
that Eqs. (\ref{eq:auxil1a}, \ref{eq:qs}) present an extension of the
well-known back-reaction boundary condition in Laplace domain (see,
e.g., \cite{Agmon90,Prustel13}) by accounting for the rate $\mu$.

The solution of the above equation can be obtained in terms of the
Legendre functions of the first kind, $P_\nu(x)$, that satisfy the
Legendre equation on $a < x < 1$,
\begin{equation}  \label{eq:eigenfunction}
\underbrace{\partial_x (1-x^2) \partial_x}_{\L} P_\nu(x) + \nu(\nu+1) P_\nu(x) = 0  ,
\end{equation}
and are regular at $x = 1$ (see Appendix \ref{sec:Legendre} for some
properties of these functions).  The Robin boundary condition at $x =
a$ reads
\begin{equation}  \label{eq:Robin_P}
P'_\nu(a) = q_s \, P_\nu(a) ,
\end{equation}
where the prime denotes the derivative with respect to $x$:
$P'_{\nu}(a) = (\partial_x P_{\nu}(x))_{x=a}$.  As $a$ and $q_s$ are
fixed, this is an equation on the degree $\nu$ that has infinitely
many real-valued solutions, denoted as $\nu_n^s$, with $n =
0,1,2,\ldots$ (see Appendix \ref{sec:spectral_lam0}):
\begin{equation}
0 \leq \nu_0^s \leq \nu_1^s \leq \ldots ~  \nearrow +\infty
\end{equation}
(negative solutions are not considered because of the symmetry:
$P_\nu(x) = P_{-\nu-1}(x)$).  The eigenvalues of the operator $-\L$
are then $\nu_n^s (\nu_n^s+1)$.  We emphasize that $\nu_n^s$ and thus
the eigenfunctions $P_{\nu_n^s}(x)$ depend on $s$ through the
parameter $q_s$ in the Robin boundary condition (\ref{eq:Robin_P}).

As the operator $\L$ is self-adjoint, its eigenfunctions
$P_{\nu_n^s}(x)$ form a complete basis in the space of
square-integrable functions on $(a,1)$.  As a consequence, one can
search for a solution of Eq. (\ref{eq:eigen_prob}) in the form
\begin{equation}  \label{eq:anzats}
\tilde c(x,s) = \sum\limits_{n=0}^\infty a_n(s) \, b_n \, P_{\nu_n^s}(x) ,
\end{equation}
where the constants $b_n$ ensure the $L_2$-normalization of
$P_{\nu_n^s}(x)$,
\begin{equation}  \label{eq:cn_def}
b_n = \left(\int\limits_{a}^1 dx \, [P_{\nu_n^s}(x)]^2 \right)^{-1/2}
\end{equation}
(see the identity (\ref{eq:cn_exact}) for computing these constants).
The unknown coefficients $a_n(s)$ can be found by substituting the
anzats (\ref{eq:anzats}) into Eq. (\ref{eq:eigen_prob}), multiplying
by $P_{\nu_m^s}(x)$, integrating from $a$ to $1$, and using the
orthogonality of $P_{\nu_n^s}(x)$ (see Appendix
\ref{sec:spectral_lam0}):
\begin{equation} \label{eq:ans}
a_n(s) = \frac{C_0 \, b_n}{s + D \nu_n^s(\nu_n^s+1)/R^2} \int\limits_a^1 dx \, P_{\nu_n^s}(x) \,.
\end{equation} 
The integral of the Legendre function can be easily calculated via
Eq. (\ref{eq:Pint}), yielding
\begin{equation}  \label{eq:tildev}
\tilde c(x,s) = C_0 (1-a^2) \sum\limits_{n=0}^\infty \frac{b_n^2 \, P_{\nu_n^s}(x) \, P'_{\nu_n^s}(a)}
{\nu_n^s(\nu_n^s+1)(s + D\nu_n^s(\nu_n^s+1)/R^2)} \, .
\end{equation}
While we used here the uniform initial condition for $c(x,t)$, one can
substitute it by any given (axially symmetric) initial condition,
including the Dirac distribution $\delta(x-x_0)$, in which case the
integral of $P_{\nu_n^s}(x)$ in Eq. (\ref{eq:ans}) would simply yield
$P_{\nu_n^s}(x_0)$.

From Eqs. (\ref{eq:Jt_def}, \ref{eq:tildev}), one deduces the Laplace
transform of the total diffusive flux onto the target:
\begin{equation}   \label{eq:tildeJ}
\tilde J(s) = 2\pi C_0 D (1 - a^2)^2  \sum\limits_{n=0}^\infty \frac{b_n^2 \, [P'_{\nu_n^s}(a)]^2}
{\nu_n^s(\nu_n^s+1)\bigl(s + \frac{D}{R^2}\nu_n^s(\nu_n^s+1)\bigr)}  \,. 
\end{equation}
Eqs. (\ref{eq:tildev}, \ref{eq:tildeJ}) constitute the exact solution
of the problem in Laplace domain.

\subsection{Solution in time domain}

The Laplace transform inversion can be performed with the aid of the
residue theorem that requires finding the poles of
Eq. (\ref{eq:tildev}) in the complex plane of $s$.  We recall that
$\nu_n^s$ and thus $b_n$ and $P_{\nu_n^s}(x)$ depend on $s$ through
the parameter $q_s$ in the Robin boundary condition
(\ref{eq:Robin_P}).  Since we deal with a diffusion problem, the poles
should lie exclusively on the negative real axis.  We expect that
$P_{\nu_n^s}(x)$ as a function of $s$ does not have poles (see a short
comment in Appendix \ref{sec:spectral_lam0}, even though a rigorous
proof of this claim is beyond the scope of this paper) so that the
poles come from the zeros of the denominator in Eq. (\ref{eq:tildev}).
In the particular case $\lambda > 0$ and $\mu = 0$, the solution
reaches a nontrivial steady-state limit with $\nu = 0$ which is
determined by the pole $s = 0$ that we will treat separately (in other
cases, this pole does not appear).  The other poles are strictly
negative and determined as the zeros of the functions
\begin{equation}  \label{eq:poles}
f_n(s) \equiv s + \frac{D}{R^2} \nu_n^s(\nu_n^s + 1) = 0.
\end{equation}
Since $\nu_n^s$ as solutions of Eq. (\ref{eq:Robin_P}) depend on $s$,
one has to search for a pair $(s,\nu)$ as a solution of the system of
two nonlinear equations (\ref{eq:Robin_P}, \ref{eq:poles}).
Expressing $s$ from Eq. (\ref{eq:poles}) in terms of $\nu_n^s$, one
reduces this system to a single equation
\begin{equation} \label{eq:Robin_P2}
(\llambda + \mmu - \nu(\nu+1)) P'_\nu(a) = q(\mmu - \nu(\nu+1)) P_\nu(a) ,
\end{equation}
with dimensionless rates
\begin{equation}
\llambda = \lambda R^2/D,  \qquad  \quad \mmu = \mu R^2/D ,
\end{equation}
and $q$ is given by Eq. (\ref{eq:q}).  In Appendix
\ref{sec:spectral_lam}, we prove that all the solutions of this
equation are real; moreover, negative solutions are ignored due to the
symmetry $P_\nu(x) = P_{-\nu-1}(x)$.  We denote the positive solutions
as
\begin{equation}
0 \leq \nu_0 \leq \nu_1 \leq \ldots ~ \nearrow +\infty
\end{equation}
(without the superscript $s$) to distinguish them from $\nu_n^s$.  The
poles are simply
\begin{equation}  \label{eq:sn}
s_n = - D \nu_n(\nu_n+1)/R^2.
\end{equation}
As $\nu_n$ obeys the boundary condition (\ref{eq:Robin_P}) for such
$s$ that satisfies Eq. (\ref{eq:poles}), one can formally write $\nu_n
= \nu_n^{s_n}$.

Quite unusually, the numerical computation in time domain turns out to
be simpler than that in the Laplace domain, because the set of
solutions of Eq. (\ref{eq:Robin_P2}) has to be computed only once,
whereas solutions of Eq. (\ref{eq:Robin_P}) had to be found for each
value of $s$.

If all poles $s_n$ are simple, then the inverse Laplace transform of
the concentration reads
\begin{align}  \label{eq:vt}
c(x,t) & = C_0 \sum\limits_{n=0}^\infty c_n \, e^{- Dt \nu_n(\nu_n+1)/R^2} \, P_{\nu_n}(x) , \\  \label{eq:cn}
c_n & = (1-a^2) \frac{b_n^2  \, P'_{\nu_n}(a)}{\nu_n(\nu_n+1) (\partial_s f_n)(s_n)} \, ,
\end{align}
and that of the total flux is
\begin{align}  \label{eq:Jt}  
J(t) & = C_0 D \sum\limits_{n=0}^\infty J_n \, e^{- Dt\nu_n(\nu_n+1)/R^2} \,,  \\  \label{eq:Jn}
J_n & = 2\pi (1-a^2)^2 \, \frac{b_n^2 \, [P'_{\nu_n}(a)]^2 }{\nu_n(\nu_n+1) (\partial_s f_n)(s_n)} \,.
\end{align}
In Appendix \ref{sec:residues}, we show how the contribution
$(\partial_s f_n)(s_n)$ to each residue can be computed in practice,
whereas Appendix \ref{sec:Anumerics} presents an example of such
computation.

Finally, we also get from Eqs. (\ref{eq:nT}, \ref{eq:nC},
\ref{eq:Jt}):
\begin{equation}  \label{eq:nT_final}
n_T(t) = C_0 D \sum\limits_{n=0}^\infty J_n \, \frac{e^{-\mu t} - e^{-Dt\nu_n(\nu_n+1)/R^2}}{D\nu_n(\nu_n+1)/R^2 - \mu}
\end{equation}
(note that if $\mu$ coincides with $D\nu_n(\nu_n+1)/R^2$ for some $n$,
the corresponding term would be $J_n \, t\, e^{-\mu t}$; similarly, a
specific term appears in the case when $\mu = 0$ and $\nu_0 = 0$) and
\begin{align}
n_C(t) & = \mu C_0 D \sum\limits_{n=0}^\infty \frac{J_n}{D\nu_n(\nu_n+1)/R^2 - \mu} \\  \nonumber
& \times \biggl(\frac{1-e^{-\mu t}}{\mu} - \frac{1-e^{-Dt\nu_n(\nu_n+1)/R^2}}{D\nu_n(\nu_n+1)/R^2}\biggr)
\end{align}
(again, a specific term appears if $\mu$ coincides with
$D\nu_n(\nu_n+1)/R^2$ for some $n$).  With these expressions, we
completed obtaining the exact solution of the our model of
diffusion-influenced reactions on the sphere.

\section{Discussion}
\label{sec:discussion}

In this section, we discuss the main features of the derived
solutions.

\subsection{Conservation of particles}

When $\mu = 0$, the total number of particles $B$ is preserved.  In
fact, integrating Eq. (\ref{eq:v_diff}) over the domain $\Omega$, we
get an equation for the total number of particles diffusing on the
sphere:
\begin{equation}
\partial_t N(t) = D \int\limits_\Omega ds \, \Delta c =  \int\limits_\pa dl \, (D \partial_n c)_{|\pa} = - \partial_t n_T(t),
\end{equation}
with the last equality coming from Eq. (\ref{eq:vT_eq}).
The total number of particles, $N(t) + n_T(t)$, is thus preserved:
\begin{equation}  \label{eq:Ntotal}
N(t) + n_T(t) = N(0) + n_T(0) = 2\pi (1-a) R^2 C_0 ,
\end{equation}
where we used the initial conditions $c(x,t=0) = C_0$ and $n_T(t=0) =
0$.

In the steady-state limit (as $t\to\infty$), the populations of
particles on the surface and on the target are equilibrated.  As the
steady-state flux is zero, Eq. (\ref{eq:v_BC}) implies
\begin{equation}  \label{eq:cinf}
c_\infty = c(x,t=\infty) = \frac{\lambda}{2\pi \rho\kappa}\, n_T(t=\infty) , 
\end{equation}
from which Eq. (\ref{eq:Ntotal}) yields
\begin{equation} \label{eq:v_asympt}
c_\infty = C_0 \biggl(1 + \frac{\kappa}{\lambda R} \sqrt{\frac{1+a}{1-a}}\biggr)^{-1} \,.
\end{equation}
This expression results from the detailed balance between free and
bound particles $B$, which is controlled by the reactivity $\kappa$
and the dissociation rate $\lambda$.

\subsection{Survival probabilities}

While the above discussion dealt with the concentration $c(x,t)$ of
particles on the sphere, it is instructive to provide its
probabilistic interpretation.  Given that the initial concentration
was set to be a constant $C_0$, the dimensionless ratio $c(x,t)/C_0$
can be interpreted as the probability of finding a particle $B$, that
started from a point $x = \cos \theta$ at time $0$, in the free state
(not bound to the target) at a later time $t$.  For irreversible
reactions, it is called the survival probability \cite{Redner} because
irreversible binding to the target can be understood as ``killing'' of
a free particle $B$.  The same term was adopted by Agmon {\it et al.}
for reversible reactions in \cite{Agmon84,Agmon90,Prustel13}, in which
$c(x,t)/C_0$ was denoted as $S_{\rm rev}(t|x)$.

Moreover, Agmon {\it et al.} also considered the probability $S_{\rm
rev}(t|*)$ for a particle, that was initially in the bound state, to
be unbound in a later time $t$.  This probability was expressed in
terms of $S_{\rm rev}(t|a)$ (see \cite{Agmon90}, Eq. (3.15)).
Comparing that relation with Eq. (\ref{eq:auxil2}) at $\mu = 0$, one
can identify
\begin{equation}  \label{eq:Srev*def}
S_{\rm rev}(t|*) = \frac{\lambda}{2\pi \rho \kappa C_0} \, n_T(t) .
\end{equation}
The explicit solution (\ref{eq:nT_final}) yields thus
\begin{equation}  \label{eq:Srev*}
S_{\rm rev}(t|*) = \frac{\lambda R^2}{2\pi \rho \kappa}
\sum\limits_{n=0}^\infty J_n \, \frac{1 - e^{-Dt\nu_n(\nu_n+1)/R^2}}{\nu_n(\nu_n+1)} \,.
\end{equation}
In the long-time limit, this probability approaches a steady-state
limit, which can be either obtained from the above spectral
decomposition, or from Eqs. (\ref{eq:cinf}, \ref{eq:v_asympt},
\ref{eq:Srev*def}):
\begin{equation}
S_{\rm rev}(t=\infty|*) = \biggl(1 + \frac{\kappa}{\lambda R} \sqrt{\frac{1+a}{1-a}}\biggr)^{-1} \,.
\end{equation}
This probability is smaller than $1$, in contrast to the case of
diffusion in unbounded domains outside a disk or a sphere, studied in
\cite{Agmon90,Prustel12,Prustel13b}, for which $S_{\rm
rev}(t=\infty|*) = 1$.

We also note that Eq. (\ref{eq:vT_eq}) with $\mu = 0$ implies that the
time derivative of $S_{\rm rev}(t|*)$ gives back the total flux
$J(t)$, so that one recovers the relation (3.17) from \cite{Agmon90}
for the time-dependent rate coefficient:
\begin{equation}
k_{\rm rev}(t) = \frac{J(t)}{C_0} = \frac{\partial_t n_T(t)}{C_0} = \frac{2\pi \rho \kappa}{\lambda}\, \partial_t S_{\rm rev}(t|*).
\end{equation}
In the following, we focus on the concentration $c(x,t)$ and the total
flux $J(t)$, bearing in mind that other important quantities can be
directly accessed.

\subsection{Limiting cases}
\label{sec:limiting}

Our solution generalizes previous works on diffusion-influenced
reactions on the spherical surface.  We discuss several limiting cases
which yield simpler formulas.

\subsubsection*{Irreversible reactions}

For irreversible reactions ($\lambda = 0$), a particle $B$ that is
bound to the target, is never released.  As a consequence, the
boundary condition (\ref{eq:v_BC}) for the concentration $c$ is
decoupled from $n_T$, so that one can solve first the boundary value
problem for $c$ with Robin boundary condition, evaluate the diffusive
flux $J(t)$, and then determine $n_T$ from Eq. (\ref{eq:vT_eq}).  In
fact, Eq. (\ref{eq:qs}) is reduced to $q_s = q$, so that Robin
boundary condition (\ref{eq:Robin_P}) reads
\begin{equation} \label{eq:Robin_P1b}
P'_\nu(a) = q P_\nu(a).
\end{equation}
Its solutions $\nu_n^s$ do not depend on $s$ and actually coincide
with the solutions of $\nu_n$ of Eq. (\ref{eq:Robin_P2}).  As a
consequence, 
\begin{equation}  \label{eq:df}
(\partial_s f_n)(s_n) = 1, 
\end{equation}
and Eqs. (\ref{eq:cn}, \ref{eq:Jn}) are simplified.  In the following,
we refer to the corresponding concentration and total flux as
$c_0(x,t)$ and $J_0(t)$, where subscript $0$ highlights that this
solution corresponds to $\lambda = 0$.  This is a generalization of
the solution presented in Ref. \cite{Chao81} to the case of a
partially absorbing sink.

Whenever $a < 1$ and $\kappa > 0$ (i.e., a nontrivial partially
reactive target is present), it is easy to check that the first
solution $\nu_0$ of Eq. (\ref{eq:Robin_P1b}) is strictly positive (in
fact, if $\nu_0$ was equal to $0$, then $P_{\nu_0}(x) = 1$ and thus
Eq. (\ref{eq:Robin_P1b}) cannot be satisfied).  As a consequence, the
concentration of particles $c_0(x,t)$ and the flux $J_0(t)$ vanish as
$t\to \infty$.

\subsubsection*{Perfect sink}

For a perfect sink ($\kappa = \infty$), the Robin boundary condition
(\ref{eq:Robin_P}) is reduced to the Dirichlet condition (also known
as totally absorbing or Smoluchowski boundary condition) $c|_\pa = 0$
that reads as
\begin{equation} \label{eq:nun_Dir}
P_{\nu}(a) = 0 ,
\end{equation}
which does not depend on $s$.  As a consequence, Eq. (\ref{eq:df})
holds again, and the concentration $c(x,t)$ and the diffusive flux
$J(t)$ are still determined by Eqs. (\ref{eq:vt}, \ref{eq:Jt}), in
which $\nu_n$ are the solutions of Eq. (\ref{eq:nun_Dir}).  This is
precisely the solution presented in Ref. \cite{Chao81}.  Note that
when the sink is perfect, it does not matter whether the reaction is
reversible or not: any released particle will be immediately
re-adsorbed by the target.

\subsubsection*{Steady-state regime}

For reversible reactions with $\mu = 0$, a steady-state solution is
established at long times so that there should exist an eigenmode with
$\nu = 0$.  First, using the properties of Legendre functions, we
rewrite (\ref{eq:Robin_P}) as
\begin{equation}  \label{eq:Robin_P1a}
\bigl[a (\nu+1) - (1-a^2) q_s \bigr] P_\nu(a) = (\nu+1) P_{\nu+1}(a) .
\end{equation}
Searching for $\nu_0$ in the form $\eta s^\alpha$ (with some constant
$\eta$ and degree $\alpha$) and substituting it into
Eq. (\ref{eq:Robin_P1a}), we get with the aid of
Eqs. (\ref{eq:Pnu_eps})
\begin{equation}
\nu_0 \simeq \eta  \, s + O(s^2),  \quad \textrm{with} ~~ \eta = \frac{\kappa R}{D\lambda} \sqrt{\frac{1+a}{1-a}} \,,
\end{equation}
from which $P_{\nu_0}(x) \simeq 1 + O(s)$, $P'_{\nu_0}(a) = q_s +
O(s^2)$, $b_n^2 \simeq (1-a)^{-1}$, and thus the leading term of
Eq. (\ref{eq:tildev}) behaves asymptotically as
\begin{equation}  \label{eq:tildev_asympt}
\tilde c(x,s) \simeq  \frac{C_0}{(1 + D\eta/R^2)s} \qquad (s\to 0)\,.
\end{equation}
Inverting the Laplace transform, we retrieve a uniform steady-state
concentration in Eq. (\ref{eq:v_asympt}).

\subsection{Numerical implementation}
\label{sec:numerics}

The concentration $c(x,t)$ and the total flux $J(t)$ are given via
exact spectral decompositions (\ref{eq:vt}, \ref{eq:Jt}).  Since the
solutions $\nu_n$ grow up to infinity as $n\to\infty$, both
decompositions converge exponentially fast for any fixed $t > 0$.  As
time $t$ stands in front of $\nu_n(\nu_n+1)$ in the exponential
function, larger $Dt/R^2$ yields faster convergence.  In turn, the
effect of other parameters on the convergence speed is more subtle and
requires a systematic analysis that is beyond the scope of this paper.

In practice, the spectral decompositions have to be truncated up to
some order $\nmax$, which can be selected according to the minimal
time at which $c(x,t)$ and $J(t)$ need to be calculated.  The
properties of the Legendre function $P_\nu(x)$ summarized in Appendix
\ref{sec:Legendre} were thoroughly used for efficient numerical
computations.  In particular, $P_\nu(x)$ and its derivatives with
respect to $x$ and $\nu$ were accurately and rapidly computed via
Eqs. (\ref{eq:Pnu_int}, \ref{eq:Pnu_prime}, \ref{eq:dPnu_int}).  Using
bisection or Newton's method, one computes numerically the needed
number of the solutions $\nu_n$ of Eq. (\ref{eq:Robin_P2}).  These
solutions then determine the poles $s_n$, the normalization constants
$b_n$ and the contribution $(\partial_s f_n)(s_n)$ to each residue
(see Appendix \ref{sec:residues} for details), from which the
coefficients $c_n$ and $J_n$ are computed via Eqs. (\ref{eq:cn},
\ref{eq:Jn}).  Note that if some poles are not simple, their
contribution to the spectral decompositions (\ref{eq:vt}, \ref{eq:Jt})
should be modified according the residue theorem; in practice, the
poles were simple in all considered examples.  In Appendix
\ref{sec:Anumerics}, we illustrate this numerical scheme by computing
the total flux.

\subsection{Comparison with Tachiya's approach}
\label{sec:Tachiya}

In his seminal work \cite{Tachiya80}, Tachiya proposed an elegant
approach to relate the solutions for reversible and irreversible
reactions (see also \cite{Agmon90,Prustel13}).  His straightforward
probabilistic arguments couple $c(x,t)$ for reversible reactions (with
$\lambda > 0$) to $c_0(x,t)$ for irreversible reactions (with $\lambda
= 0$) via convolution relations that can be easily solved in the
Laplace domain.  In our notations, one gets
\begin{equation}  \label{eq:Tachiya}
\tilde{c}(x,s) = \frac{1}{s} + \frac{\tilde{c}_0(x,s) - 1/s}{1 + \lambda \, \tilde{c}_0(a,s) \, s/(s+\mu)} \,,
\end{equation}
where we also included the rate $\mu$.  One can easily check that the
function $\tilde{c}(x,s)$ satisfies Eqs. (\ref{eq:auxil1},
\ref{eq:eigen_prob}) with $q_s$ from Eq. (\ref{eq:qs}), given that
$\tilde{c}_0(x,s)$ satisfies Eqs. (\ref{eq:auxil1},
\ref{eq:eigen_prob}) with $q$ instead of $q_s$.  This is a significant
simplification because Eq. (\ref{eq:tildev}) for $\tilde c_0(x,s)$
involves $\nu_n^s$ that do not depend on $s$ and thus have to be
computed only once.  However, the simple and instructive relation
(\ref{eq:Tachiya}) in the Laplace domain is not much helpful for the
Laplace inversion.  In contrast, our approach allows one to get the
exact solutions in time domain so that two approaches are
complementary to each other.

We also note that our description with $\mu = 0$ is equivalent to that
given by Pr\"ustel and Tachiya \cite{Prustel13}.  In fact, the
boundary condition (\ref{eq:v_BC}) can be re-written in the form of
Eq. (16) from Ref. \cite{Prustel13}:
\begin{equation}
2\pi \rho \frac{D}{R} \sin\ve \partial_x c_{|\pa} = k_a c_{|\pa} - k_d \biggl(1 - 2\pi R^2 \int\limits_{a}^1 dx' \, c(x',t)\biggr) ,
\end{equation}
with $k_a = 2\pi \rho \kappa$, $k_d = \lambda$, and
\begin{equation}
n_T(t) = 1 - 2\pi R^2 \int\limits_{a}^1 dx' \, c(x',t).
\end{equation}
The time derivative of the last equation yields, after
simplifications, Eq. (\ref{eq:vT_eq}).

Instead of solving exactly the problem for irreversible kinetics,
Pr\"ustel and Tachiya used an exponential approximation for the
survival probability,
\begin{equation}  \label{eq:vt_irrev_approx}
S_0(x,t) = \frac{c_0(x,t)}{C_0} \approx \exp(- t/ \tau(x)), 
\end{equation}
where $\tau(x)$ is the mean first passage time in the case of
irreversible reaction (with $\lambda = 0$):
\begin{equation}
\tau(x) = \int\limits_0^\infty dt\, S_0(x,t) \,,
\end{equation}
which was found in \cite{Sano81} (see also Appendix
\ref{sec:MFPT_Robin}):
\begin{equation}  \label{eq:MFPT_Robin}
\tau(x) = \frac{R^2}{D} \ln \biggl(\frac{1+x}{1+a}\biggr) + \frac{R}{\kappa} \sqrt{\frac{1-a}{1+a}} \, .
\end{equation}
With this approximation, the Laplace inversion of
Eq. (\ref{eq:Tachiya}) yielded an approximate formula for the
concentration $c(x,t)$ which reads in our notations as
\begin{eqnarray}  \nonumber  
&& \frac{c(x,t)}{C_0} \simeq  \frac{\lambda}{\lambda + \frac{1}{\tau(a)}} 
+ e^{-t/\tau(x)} \frac{\frac{1}{\tau(x)} - \frac{1}{\tau(a)}}{\frac{1}{\tau(x)} - \frac{1}{\tau(a)} - \lambda} \\
\label{eq:vt_approx}
&& - e^{-t(\lambda + 1/\tau(a))} \biggl[\frac{\lambda}{\lambda + \frac{1}{\tau(a)}} 
+ \frac{\lambda}{\frac{1}{\tau(x)} - \frac{1}{\tau(a)} - \lambda} \biggr]  . 
\end{eqnarray}

\begin{figure}
\begin{center}
\includegraphics[width=88mm]{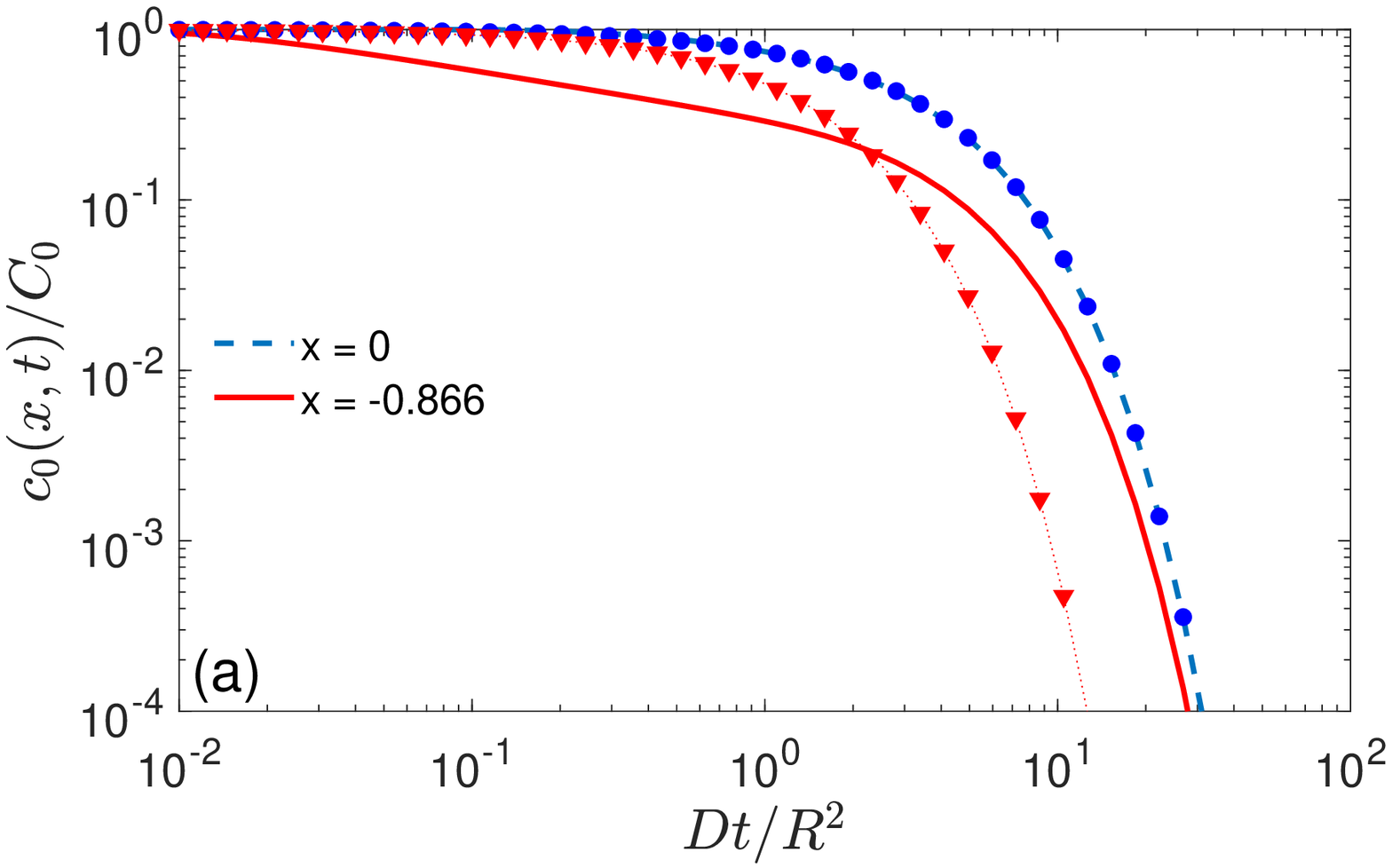} 
\includegraphics[width=88mm]{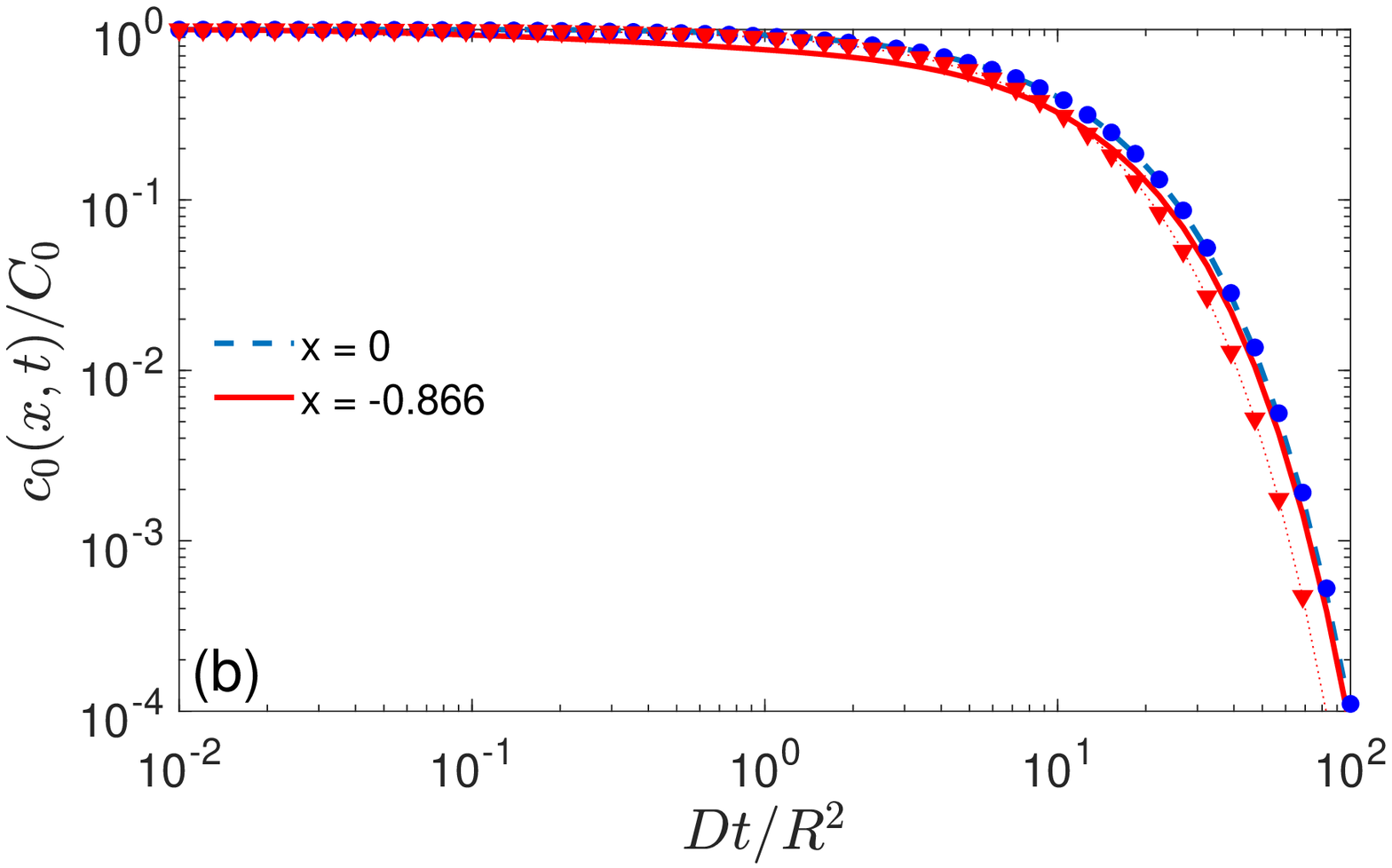} 
\end{center}
\caption{
The normalized concentration $c_0(x,t)/C_0$ for irreversible reactions
($\lambda = 0$) as a function of rescaled time $Dt/R^2$, for a target
of angular size $\ve = \pi/12$, and two starting points: $x = 0$ (the
equator) and $x = \cos(\pi - 2\ve) \approx -0.886$ (close to the
target).  Lines show the exact solution (\ref{eq:vt}), truncated at
$\nmax = 50$, while symbols present the exponential approximation
(\ref{eq:vt_irrev_approx}).  {\bf (a)} Perfectly reactive target
($\kappa = \infty$); {\bf (b)} partially reactive target ($\kappa R/D
= 1$).}
\label{fig:vt0}
\end{figure}

Since the validity of this approximation was not tested in
\cite{Prustel13}, we undertake this study here.  Figure \ref{fig:vt0}
illustrates the quality of the exponential approximation
(\ref{eq:vt_irrev_approx}) for irreversible reactions.  When the
particle $B$ starts far from a small perfectly absorbing target ($x =
0$, $\kappa = \infty$, Fig. \ref{fig:vt0}a), the diffusive search
takes long time that is enough for a particle to ``forget'' about its
starting point.  In this setting, Eq.  (\ref{eq:vt_irrev_approx})
accurately approximates the survival probability.  In contrast, when
the starting point is close to the target ($x = -\cos(2\ve) \approx
-0.866$), there are high changes of finding it in a short time, and
the exponential approximation (\ref{eq:vt_irrev_approx}) fails.  It is
also worth noting that the exponential asymptotic decay of the
survival probability $S_0(x,t)$ is determined by the smallest
eigenvalue of the diffusion operator, which is independent of the
starting point.  In this light, the approximation
(\ref{eq:vt_irrev_approx}) can only be valid when $\tau(x)$ is close
to $T = R^2/(D\nu_0(\nu_0+1))$ and thus independent of $x$, requiring
the starting point to be far from the target and the target to be
small.  The situation is different for weakly reactive target with
$\kappa R/D = 1$ (Fig. \ref{fig:vt0}b).  Even when the starting point
is close to the target, multiple failed attempts of binding to the
target allow for the particle to explore the whole domain until the
successful binding.  In other words, the second term in
Eq. (\ref{eq:MFPT_Robin}) is dominant, whereas the dependence on the
starting point $x$ is irrelevant.  In summary, for a fixed $x$, the
approximation (\ref{eq:vt_approx}) is getting more accurate when the
target size and/or reactivity decrease.  In particular, this
approximation can be employed to re-derive the asymptotic behavior of
the total flux in the small-target limit, as an alternative to the
rigorous analysis presented in Sec. \ref{sec:small-target}.  At the
same time, the dependence on the starting point is not captured, and
the approximation fails when the starting point is close to a highly
reactive target, irrespectively of its size.  In this light, an even
simpler approximation $S_0(x,t) \approx \exp(-t/\tau(0))$ would be
more natural.

\begin{figure}
\begin{center}
\includegraphics[width=88mm]{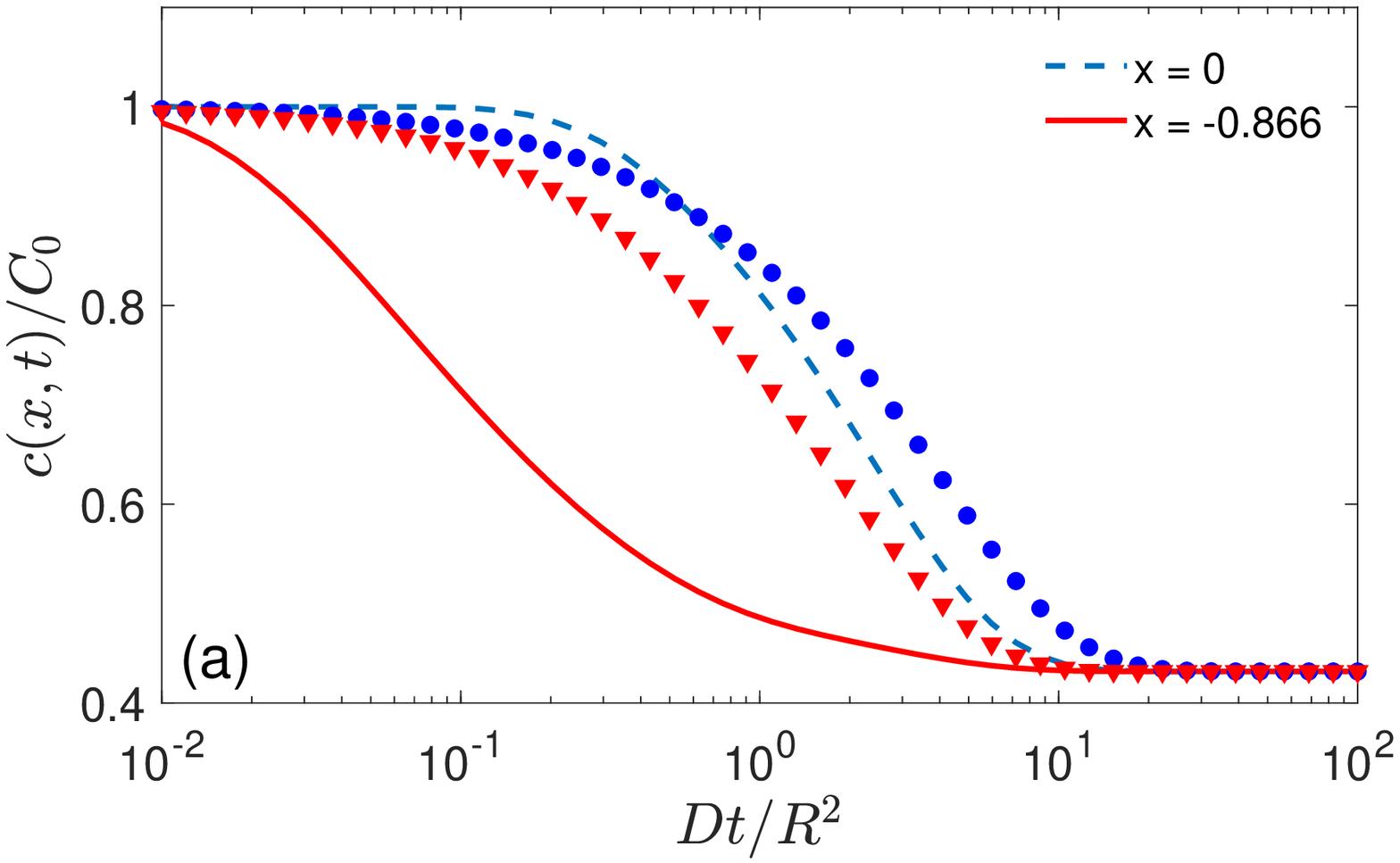} 
\includegraphics[width=88mm]{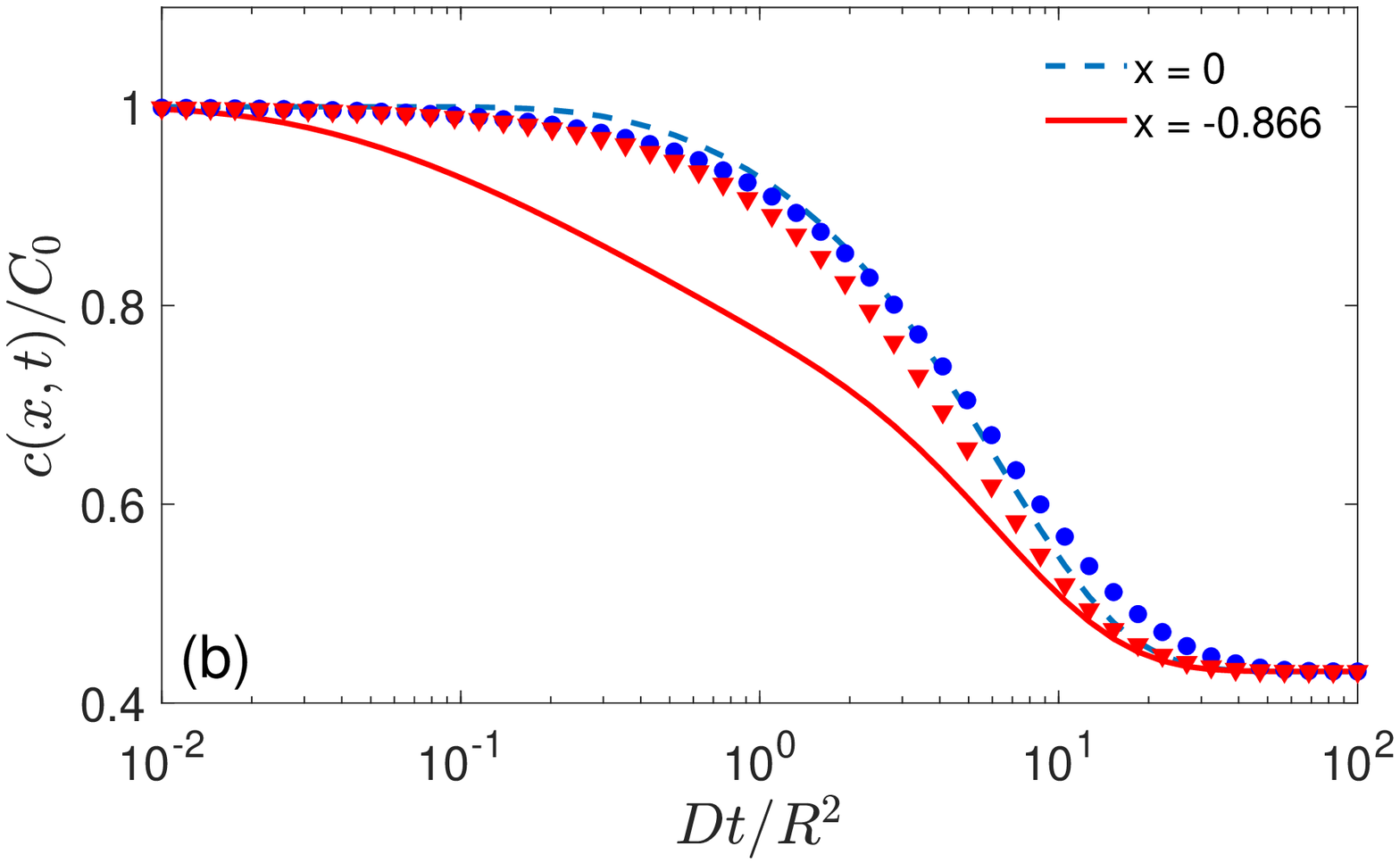} 
\end{center}
\caption{
The normalized concentration $c(x,t)/C_0$ for reversible reactions
(with $\mu = 0$) as a function of rescaled time $Dt/R^2$, for a target
of angular size $\ve = \pi/12$, and two starting points: $x = 0$ (the
equator) and $x = \cos(\pi - 2\ve) \approx -0.886$ (close to the
target).  Lines show the exact solution (\ref{eq:vt}), truncated at
$\nmax = 20$, while symbols present the approximate solution
(\ref{eq:vt_approx}).  {\bf (a)} $\kappa R/D = 10$ and $\lambda R^2/D
= 1$; {\bf (b)} $\kappa R/D = 1$ and $\lambda R^2/D = 0.1$.}
\label{fig:vt}
\end{figure}

Figure \ref{fig:vt} shows the normalized concentration $c(x,t)/C_0$
for reversible reactions (with $\lambda > 0$ and $\mu = 0$) as a
function of rescaled time $Dt/R^2$ for the same target and the same
starting points.  As a perfectly reactive target would yield
irreversible reactions, we do not consider the case $\kappa = \infty$
but compare two cases with high reactivity $\kappa R/D = 10$ and weak
reactivity $\kappa R/D = 1$.  To keep the same steady-state
concentration $c_\infty$, we set $\lambda R^2/D = 1$ and $\lambda
R^2/D = 0.1$, respectively.  As discussed previously for irreversible
reactions, the approximate solution (\ref{eq:vt_approx}) fails when
the starting point is close to the target, even for weakly reactive
target.  In turn, the approximation is getting more accurate when the
starting point is far from the target and the reactivity is low.

\subsection{Small-target asymptotic behavior}
\label{sec:small-target}

In many practical applications, the radius $\rho$ of the target is
much smaller than the radius $R$ of the sphere, i.e., $\ve \ll 1$ and
$a$ is close $-1$.  In the conventional first-passage time framework,
this situation is known as the narrow escape problem
\cite{Holcman14,Metzler}.  Here we summarize the asymptotic behavior
of our solutions in the narrow escape limit $\ve \to 0$ (see details
in Appendix \ref{sec:asympt}).

In the limit $\ve = 0$, a point-like target cannot be accessed by
Brownian motion on the sphere and thus corresponds to no target
situation, in which the spectrum of the governing operator $\L$ is
simply $\nu_n = n$ for $n = 0,1,2,\ldots$, whereas the eigenfunctions
$P_\nu(x)$ are the Legendre polynomials.  In particular, the
concentration $c(x,t)$ remains constant, whereas the flux $J(t)$ is
zero.  

When $\ve$ is small but nonzero, finding such a small target is a long
search process.  In Appendix \ref{sec:asympt}, we deduce the long-time
behavior of the total flux as
\begin{equation}  \label{eq:Jtinf}
J(t) \simeq \frac{4\pi C_0 R^2}{T} \, e^{-t/T}   \qquad (t \to \infty)\,,
\end{equation}
with a characteristic time $T$, which is determined by the smallest
eigenvalue of the diffusion operator:
\begin{equation}
T = \frac{R^2}{D \nu_0(\nu_0+1)} \simeq \frac{R^2}{D\nu_0}   \qquad (\ve \ll 1).
\end{equation}
Since the factor $N_B = 4\pi C_0 R^2 \simeq 2\pi(1-a)C_0 R^2$ is
simply the total initial number of diffusing particles $B$, the above
expression (\ref{eq:Jtinf}) can be understood as an exponential
distribution of the first-passage times for $N_B$ independent
particles.  The asymptotic behavior of the smallest degree $\nu_0$ was
obtained in Appendix \ref{sec:asympt} for four different scenarios:

(i) For a perfectly reactive target ($\kappa = \infty$), one gets
\begin{equation}  \label{eq:nu0_kinf}
\nu_0 \simeq \frac{1}{2\ln(2/\ve)} \qquad (\ve\ll 1)\,.
\end{equation}
The associated time scale $T = 2R^2 \ln(2/\ve)/D$ is close to the mean
FPT to a perfectly reactive target (the first term in
Eq. (\ref{eq:MFPT_Robin})), see also \cite{Singer06c}.

(ii) For irreversible reactions ($\lambda = 0$, $\kappa < \infty$),
the smallest degree $\nu_0$ vanishes linearly with $\ve$, 
\begin{equation}  \label{eq:nu0_lambda0}
\nu_0 \simeq \frac{\kappa \ve R}{2D} \approx \frac{\kappa \rho}{2D}  \qquad (\ve\ll 1).
\end{equation}
The associated time scale $T = 2R^2/(\kappa \rho)$ does not depend on
the diffusion coefficient and is close to the mean FPT to a weakly
reactive target (the second term in Eq. (\ref{eq:MFPT_Robin})), see
also \cite{Coombs09}.

(iii) For reversible reactions ($\lambda > 0$, $\mu = 0$), the
smallest degree $\nu_0$ is equal to $0$ for any target size.  In this
case, the steady-state total flux is zero, $J(\infty) = 0$, while the
steady-state concentration remains close to the initial concentration
$C_0$:
\begin{equation}
c_\infty \simeq C_0 \biggl(1 + \frac{\kappa \rho}{2\lambda}\biggr)^{-1} ,
\end{equation}
as expected.  The approach to the steady state is controlled by the
next eigenvalue, $\nu_1(\nu_1+1) = 2 + O(\ve)$, which sets the time
scale $R^2/(2D)$.  However, if the dissociation rate $\lambda$ is
small, $\llambda < 1$, the approach is controlled by the solution in
Eq. (\ref{eq:nu_special}), which sets the longer time scale
$1/\lambda$.

(iv) For partly reversible reactions ($\lambda > 0$, $\mu > 0$), we
obtain (see Appendix \ref{sec:asympt})
\begin{equation}  \label{eq:nu0_general}
\nu_0 \simeq \frac{\mu \kappa \rho}{2D(\mu+\lambda)} \,,
\end{equation}
which is reduced to the second situation by setting $\lambda = 0$ and
to the third situation by setting $\mu = 0$.  This value determines
the time scale $T = 2R^2(1 + \lambda/\mu)/(\kappa \rho)$.

In analogy to the exact solution (\ref{eq:MFPT_Robin}) for the mean
FPT, we propose an interpolation formula for the time scale $T$:
\begin{equation}  \label{eq:Tinterp}
T = \frac{2R^2 \ln(2R/\rho)}{D} + \frac{2R^2(1 + \lambda/\mu)}{\kappa \rho} \,.
\end{equation}
In the limit $\kappa = \infty$, one retrieves the time scale for a
perfectly reactive target.  In turn, for moderate values of $\kappa$,
the second term provides the dominant contribution in the small-target
limit ($\rho \ll R$).

\begin{figure}
\begin{center}
\includegraphics[width=88mm]{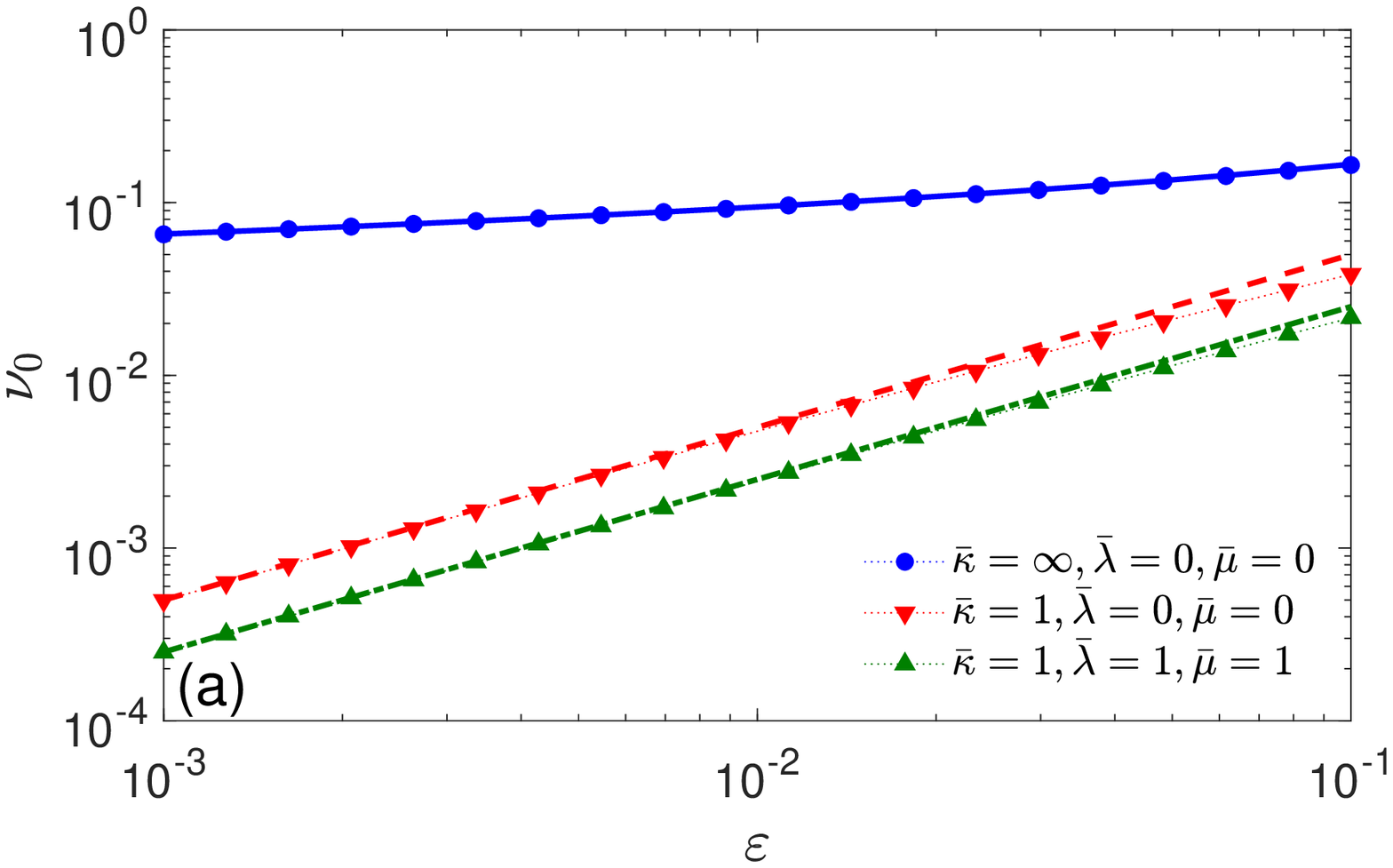} 
\includegraphics[width=88mm]{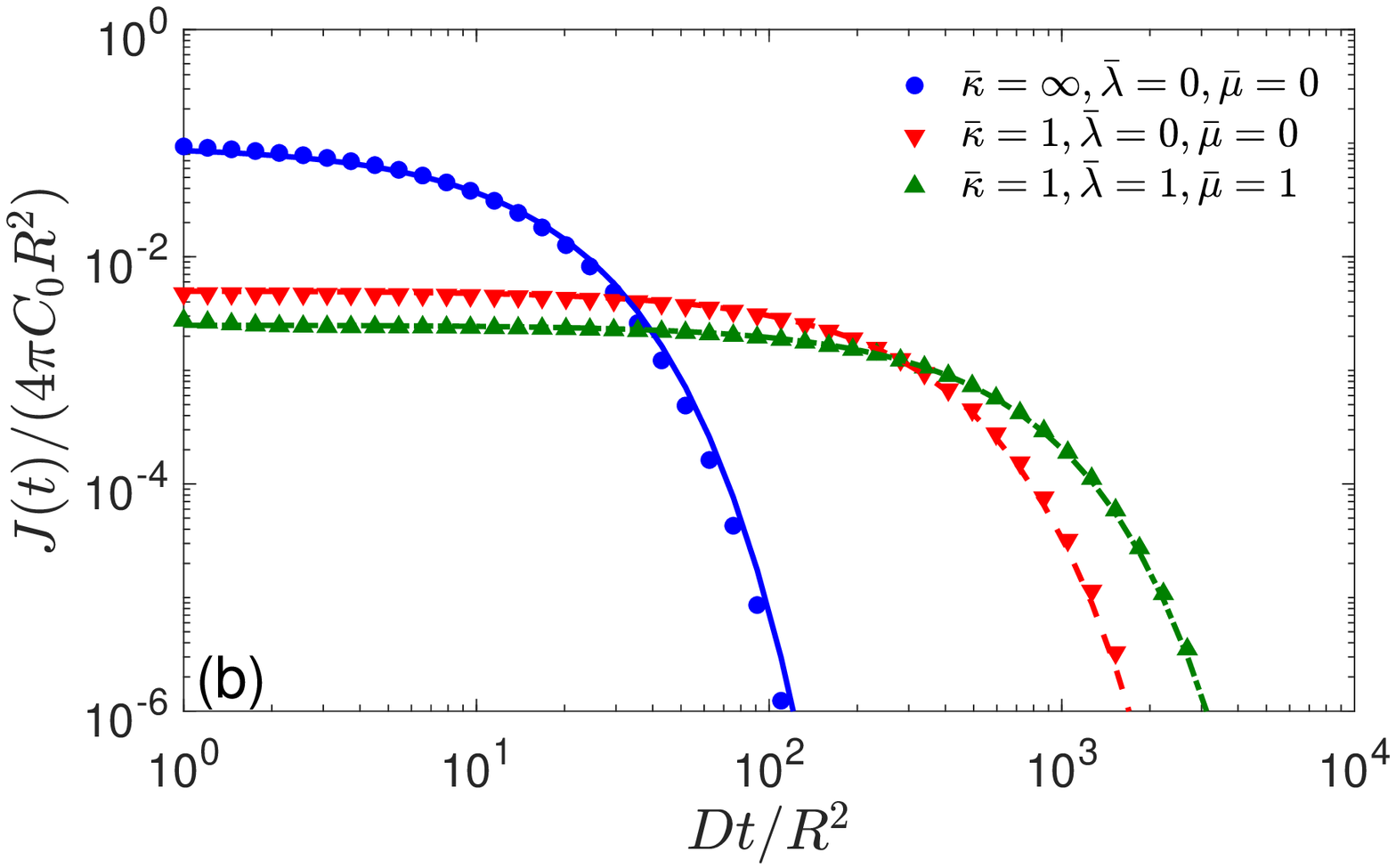} 
\end{center}
\caption{
{\bf (a)} The smallest degree $\nu_0$ as a function of the target
angular size $\ve$, for three configurations of reaction parameters:
perfectly reactive target ($\bar{\kappa} = \infty$), partially
reactive target with irreversible binding ($\bar{\kappa} = 1$,
$\bar{\lambda} = 0$), and partially reactive target with partly
reversible binding ($\bar{\kappa} = 1$, $\bar{\lambda} = 1$,
$\bar{\mu} = 1$).  Symbols show the numerical solutions of
Eq. (\ref{eq:Robin_P1b}) for the first two cases, and of
Eq. (\ref{eq:Robin_P2}) for the last case.  Lines show the asymptotic
results (\ref{eq:nu0_kinf}, \ref{eq:nu0_lambda0},
\ref{eq:nu0_general}).  {\bf (b)} The total flux $J(t)$ (normalized by
$4\pi C_0 R^2$) as a function of rescaled time $Dt/R^2$, for a target
of the angular size $\ve = 0.01$, and three configurations of reaction
parameters as above.  Symbols show the exact solution (\ref{eq:Jt}),
truncated at $\nmax = 20$, while lines present the long-time
asymptotic formula (\ref{eq:Jtinf}). }
\label{fig:nu0}
\end{figure}

Figure \ref{fig:nu0}(a) illustrates an excellent agreement between
numerically computed values of $\nu_0$ and its asymptotic relations.
In addition, Fig. \ref{fig:nu0}(b) shows the accuracy of the long-time
approximation (\ref{eq:Jtinf}) for a target of angular size $\ve =
0.01$.  One can see that our explicit formulas accurately describe the
long-time behavior of the total flux toward the target in different
regimes.

\subsection{Inclusion of constant flux from the bulk}
\label{sec:bulk_flux}

Our description can be further extended e.g., by incorporating a
constant flux of particles from the bulk.  In this case, the
concentration is constantly fed, and the diffusion equation
(\ref{eq:v_diff}) can be modified by adding the constant flux density
$J$ (in units mol/m$^2$/s) to the right-hand side.  In the Laplace
domain, this is equivalent to changing $C_0$ to $C_0 + J/s$ in
Eq. (\ref{eq:auxil2}), as well as in the solutions (\ref{eq:tildev},
\ref{eq:tildeJ}).  As a consequence, one gets
\begin{equation}  \label{eq:vb_t}
c_{\rm b}(x,t) = c(x,t) + \frac{J}{C_0} \int\limits_0^t dt'\, c(x,t')
\end{equation}
and 
\begin{equation}  \label{eq:Jb_t}
J_{\rm b}(t) = J(t) + \frac{J}{C_0} \int\limits_0^t dt'\, J(t'),
\end{equation}
where $c(x,t)$ and $J(t)$ are the former solutions without bulk flux.
Substitution of the exact decompositions (\ref{eq:vt},
\ref{eq:Jt}) into Eqs. (\ref{eq:vb_t}, \ref{eq:Jb_t}) yields
\begin{align}
c_{\rm b}(x,t) = c_{\rm b}(x,\infty) & + C_0\sum\limits_{n=0}^\infty \biggl[1 - \frac{J R^2}{DC_0\nu_n(\nu_n+1)}\biggr] \\  \nonumber
& \times c_n \, P_{\nu_n}(x) \, e^{-Dt \nu_n(\nu_n+1)/R^2} 
\end{align}
and
\begin{align}
J_{\rm b}(t) = J_{\rm b}(\infty) + C_0 D\sum\limits_{n=0}^\infty & \biggl[1 - \frac{J R^2}{DC_0\nu_n(\nu_n+1)}\biggr] \\  \nonumber
& \times J_n \, e^{-Dt \nu_n(\nu_n+1)/R^2} ,
\end{align}
where 
\begin{align}
c_{\rm b}(x,\infty) & = \frac{J R^2}{D} \sum\limits_{n=0}^\infty \frac{c_n}{\nu_n(\nu_n+1)} \, P_{\nu_n}(x) , \\
J_{\rm b}(\infty) & = J R^2 \sum\limits_{n=0}^\infty \frac{J_n}{\nu_n(\nu_n+1)} 
\end{align}
are the steady-state concentration and diffusive flux, respectively.
In contrast to the case without bulk flux, the concentration and the
diffusive flux do not vanish as $t\to \infty$ but reach a steady-state
regime.  Expectedly, this flux does not depend on the diffusion
coefficient $D$, except for the implicit dependence via $\nu_n$.  Note
that the term $n = 0$ should be treated separately in the case of
fully reversible reactions ($\lambda > 0$, $\mu = 0$), for which
$\nu_0 = 0$.

\section{Conclusions and perspectives}
\label{sec:conclusion}

In this paper, we investigated diffusion of particles on the surface
of a sphere toward a partially reactive target with partly reversible
binding kinetics.  In this setting, a diffusing particle binds to the
target with a certain probability and then, after some residence time
on the target, it can either unbind to resume diffusion, or be
transformed into another particle.  These reaction mechanisms were
characterized by three parameters: the reactivity $\kappa$
(controlling binding to the target), the dissociation rate $\lambda$
(controlling unbinding from the target), and the transformation rate
$\mu$ (controlling the ultimate irreversible chemical transformation).
This is a rather general reaction model that englobes most of formerly
considered situations.  In particular, we distinguished four
scenarios: (i) an ideal sink ($\kappa = \infty$; $\lambda$ does not
matter), (ii) a partially reactive target with irreversible binding
($\kappa < \infty$ and $\lambda = 0$), (iii) a partially reactive
target with reversible binding and no chemical transformation ($\kappa
< \infty$, $\lambda > 0$, $\mu = 0$), and (iv) a partially reactive
target with reversible binding and chemical transformation ($\kappa <
\infty$, $\lambda > 0$, $\mu > 0$).  Most former studies were
dedicated to the first and (less) to the second situation.

Employing the axial symmetry of the problem, we derived the exact
solution of the underlying diffusion-reaction equations in terms of
Legendre functions.  While the case of irreversible reactions on the
surface of a sphere was studied in the past \cite{Chao81,Sano81}, the
coupling of differential equations for reversible reactions presented
a mathematical obstacle in getting an exact solution in time domain.
We solved this problem and derived a new equation determining the
eigenvalues of the governing diffusion operator.  While our solution
is valid for any target size, further simplifications were achieved in
the small-target limit.  Here, we determined the asymptotic behavior
of the eigenvalues of the governing operator.  In particular, the
smallest eigenvalue controls the long-time exponential decay of the
total flux.  Moreover, we proposed a simple interpolation formula
(\ref{eq:Tinterp}) for the time scale $T$ of this decay that includes
all the reaction parameters.

While surface diffusion on a cell membrane is an emblematic problem
that inspired our work, the proposed method is not limited to cellular
biology.  For instance, one can think of animal or human migration on
Earth, including the problems of epidemic spreading, in which
first-passage time phenomena are critically relevant, even though the
dynamics of such processes is usually more complex than ordinary
diffusion.  More generally, this geometric setting allows one to study
the effect of curvature onto diffusion-reaction processes, as compared
to a flat annulus geometry on the plane.  The latter case is simpler
and was much more thoroughly studied, especially for irreversible
reactions \cite{Carslaw,Crank,Thambynayagam,Redner}.  Following the
same lines as in Sec. \ref{sec:solution}, one can derive the exact
solution describing partly reversible diffusion-influenced reactions
(see also \cite{Prustel13}, in which such a solution was derived for
the case $\mu = 0$).  A systematic analysis of the curvature effect
onto diffusion-reaction processes presents an interesting perspective
of this work.

Due to a finite surface area of the sphere and thus a finite
amount of diffusing particles, a steady-state solution of the
underlying diffusion equation in the presence of a sink is zero, so
that one has to look for a transient, time-dependent solution.
Moreover, the passage from the bounded surface on the sphere to the
whole plane does not resolve the difficulties \cite{Torney83}.  In
fact, the recurrent character of Brownian motion implies that there is
no steady-state solution in the plane with a finite concentration at
infinity.  This is in contrast to three-dimensional unbounded domains,
for which a nonzero steady-state regime can be reached.  In
particular, the reaction rate of particles diffusing in $\R^3$ toward
a spherical sink was first computed a century ago by Smoluchowski
\cite{Smoluchowski17} and then extended to various target
configurations (see
\cite{Galanti16,Traytak18,Grebenkov18e,Grebenkov19a} and references
therein).

Our exact solution for the concentration of particles revealed
limitations of some common practices.  In particular, the survival
probability of a particle diffusing in a bounded domain toward a
target is often approximated by a single exponential function (see
\cite{Sano81,Benichou10a,Benichou14} and references therein).  This is a
long-time approximation, which is accurate when the starting point is
far from the target or when it is averaged over the whole domain.  In
both cases, the particle diffuses for long time until its binding to
the target and thus can largely explore the whole bounded domain,
resulting in such an exponential law.  However, if the particle starts
relatively close to the target, it can bind to the target very
rapidly, on a time scale much shorter than that of the diffusive
exploration of the domain \cite{Godec16,Grebenkov18a,Grebenkov18b}.
As a consequence, the exponential function fails to describe the
intricate short-time behavior of the survival probability.  This is
particularly relevant for reversible reactions because the dissociated
particle is released on the target.  For instance, we showed in
Sec. \ref{sec:Tachiya} that the approximation by Pr\"ustel and Tachiya
based on the exponential approximation may be accurate for distant
starting points but fails if the starting point is close to the
target.

The developed description of partly reversible diffusion-influenced
reactions is not limited to a spherical shape.  It can be easily
adapted to general manifolds or Euclidean domains.  While the
possibility of getting the exact solution strongly relied on the
symmetries of the spherical surface, small-target asymptotic results
may potentially be derived without knowledge of an exact solution, as
it was done for ideal sinks \cite{Singer06c,Holcman14} or partially
absorbing traps on a cylindrical dendritic membrane
\cite{Bressloff08}.  Apart from considering other domains and
manifolds, the presented approach can be extended in different
directions.  First, we showed in Sec. \ref{sec:bulk_flux} how to
incorporate a constant flux density from the bulk and to get
stationary solutions.  In this way, surface diffusion toward the
target can be coupled to bulk diffusion (see also \cite{Novak07} and
references therein).  The bulk flux may account, e.g., for particles
coming from the exterior of a cell membrane or from the interior of
the cell.  The latter case can model ``recycling'' of proteins by the
cell.  Second, one can consider diffusion on a part of the sphere
delimited by two circles of latitude.  This is equivalent to solving
the diffusion-reaction equation (\ref{eq:eigen_prob}) for $\tilde
c(x,s)$ on an interval $(a,b)$.  As previously, the endpoint $a$
corresponds to the target, whereas the circle of latitude determined
by $\cos\theta = b$ can mimic a biological frontier that confines
proteins, e.g., in a lipid raft \cite{Simons97,Masson09,Lingwood10}.
Alternatively, if there are many targets that are more or less
uniformly distributed on the surface of a sphere, one can virtually
split this surface into ``zones of influence'' of each target.  Even
if the particle can cross the virtual border of each zone of
influence, such a crossing would simply mean changing the zone of
influence and can thus be modeled by reflecting boundary condition.  A
solution in such a geometric setting would involve Legendre functions
of two kinds, $P_\nu(x)$ and $Q_\nu(x)$.  The problem with multiple
small targets may also be addressed by extending the asymptotic
methods developed in \cite{Coombs09}.  Third, one can investigate
first-passage quantities describing multiple independent particles
such as, e.g., the first moment when $n$ among $N$ particles are bound
to the target.  The reversible binding makes these statistics highly
nontrivial \cite{Grebenkov17c,Lawley19}.  Finally, an extension of the
proposed formalism beyond Brownian motion presents an important
perspective.  In fact, single-particle tracking experiments on cell
membranes witness considerable deviations from ordinary diffusion
\cite{Ritchie05,Yamamoto14,He16}.  Some theoretical models of
anomalous diffusion such as continuous-time random walks or processes
with diffusing diffusivity can be implemented into our description, at
least in Laplace domain (see discussions in
\cite{Chechkin17,Lanoiselee18,Grebenkov19,Lanoiselee19} and references
therein).

\begin{acknowledgments}
The author acknowledges Prof. D. Calebiro for inspiring discussions.
\end{acknowledgments}

\appendix
\section{Some properties of Legendre functions}
\label{sec:Legendre}

The Legendre function $P_\nu(x)$ of the first kind is the solution of
the Legendre equation (\ref{eq:eigenfunction}), which is regular at $x
= 1$ and normalized such that $P_\nu(1) = 1$.  For an integer $\nu$,
$P_\nu(x)$ is the Legendre polynomial of degree $\nu$.  In general,
the Legendre function is expressed through the hypergeometric
function:
\begin{equation}
P_\nu(x) = ~_2F_1\bigl(-\nu,\nu+1; 1; (1-x)/2\bigr) ,
\end{equation}
which is defined as
\begin{equation}  \label{eq:2F1}
~_2F_1(a,b;c;z) = \sum\limits_{n=0}^\infty \frac{(a)_n \, (b)_n}{(c)_n\, n!} \, z^n ,
\end{equation}
with $(a)_n = a(a+1)\ldots (a+n-1)$ (and $(a)_0 = 1$).  For
non-integer $\nu$, the Legendre function $P_\nu(x)$ has a logarithmic
divergence at $x = - 1$.  In turn, for any fixed $x > -1$, $P_\nu(x)$
is an entire analytic function of $\nu$ because the hypergeometric
series in Eq. (\ref{eq:2F1}) converges uniformly for any finite domain
of $\nu \in \C$, see \cite{Erdelyi} (p.68) for a similar statement for
the hypergeometric function.
Many properties of the Legendre function are discussed in the classic
textbook by Hobson \cite{Hobson}, while their applications in electric
potential problems are summarized in \cite{Hall49} (see also
\cite{Smythe,Maier16}).  Here we reproduce some properties that are
relevant for our work.

Multiplying Eq. (\ref{eq:eigenfunction}) by $P_{\nu'}(x)$, subtracting its
symmetrized version and integrating over $x$ from $a$ to $1$, one
immediately deduces
\begin{eqnarray}   \label{eq:orthogonality}
&& \frac{(\nu-\nu')(\nu+\nu'+1)}{1-a^2}\int\limits_a^1 dx \, P_\nu(x) \, P_{\nu'}(x) \\   \nonumber
&& =  P_{\nu'}(a) P'_{\nu}(a) - P_{\nu}(a) P'_{\nu'}(a) ,
\end{eqnarray}
where $P'_\nu(a) = (\partial_x P_\nu(x))_{x=a}$.  In turn, the
integral of Eq. (\ref{eq:eigenfunction}) from $a$ to $1$ yields
\begin{equation}  \label{eq:Pint}
\int\limits_a^1 dx \, P_{\nu}(x) = \frac{P_{\nu-1}(a) - P_{\nu+1}(a)}{2\nu+1} =
\frac{1-a^2}{\nu(\nu+1)} P'_\nu(a),
\end{equation}
where we used the recurrence relation
\begin{equation}
(\nu+1) P_{\nu+1}(x) = (2\nu+1) x P_\nu(x) - \nu P_{\nu-1}(x) 
\end{equation}
and the expression for the derivative with respect to $x$:
\begin{equation}  \label{eq:Pnu_prime}
P'_\nu(x) = (\nu+1) \frac{P_{\nu+1}(x) - x P_\nu(x)}{x^2-1} \,.
\end{equation}

The normalization constants $b_n$ in Eq. (\ref{eq:cn_def}) can be
computed in a standard way.  On one hand, the Legendre equation
(\ref{eq:eigenfunction}) is multiplied by $\partial_\nu P_\nu(x)$ and
integrated over $x$ from $a$ to $1$.  On the other hand, the Legendre
equation is differentiated with respect to $\nu$, multiplied by
$P_\nu(x)$ and then integrated.  Taking the difference of two
equations and integrating by parts, one gets
\begin{eqnarray}  
&& \int\limits_a^1 dx\, [P_\nu(x)]^2 = \frac{1}{2\nu+1}  \\  \nonumber
&& \times \left.
\biggl((1-x^2) \bigl[\partial_\nu P_\nu(x) \, P'_\nu(x) - P_\nu(x) \, \partial_\nu P'_\nu(x) \bigr]\biggr)\right|_{x=a}^{x=1} .
\end{eqnarray}
Taking the derivative of Eq. (\ref{eq:Pnu_prime}) with respect to
$\nu$, one finds
\begin{equation} 
\partial_\nu P'_\nu(x) = \frac{P'_\nu(x)}{\nu+1} + \frac{\nu+1}{x^2-1} \bigl(\partial_\nu P_{\nu+1}(x) - x \partial_\nu P_\nu(x)\bigr) \,,
\end{equation}
from which
\begin{eqnarray*}  
&& \int\limits_a^1 dx\, [P_\nu(x)]^2 = \frac{1}{2\nu+1}  \biggl(P_\nu(x)(P_{\nu+1}(x)-xP_\nu(x)) \\  \nonumber
&&+ \left. (\nu+1) \bigl[P_\nu(x) \partial_\nu P_{\nu+1}(x) - P_{\nu+1}(x) \partial_\nu P_\nu(x)
 \bigr]\biggr) \right|_{x=a}^{x=1} .
\end{eqnarray*}
Since $P_\nu(1) = 1$, the normalization coefficients $b_n$ in
Eq. (\ref{eq:cn_def}) are obtained from
\begin{align}  \nonumber
&\int\limits_a^1 dx\, [P_\nu(x)]^2 = \frac{-1}{2\nu+1}  \biggl(P_\nu(a)(P_{\nu+1}(a)-aP_\nu(a)) \\   \label{eq:cn_exact}
&+ (\nu+1) \bigl[P_\nu(a) \partial_\nu P_{\nu+1}(a) - P_{\nu+1}(a) \partial_\nu P_\nu(a)  \bigr]\biggr)  .
\end{align}

For Dirichlet boundary condition $P_{\nu}(a) = 0$, one recovers the
relation from \cite{Hall49}:
\begin{equation}
\int\limits_a^1 dx [P_\nu(x)]^2 = -\frac{1-a^2}{2\nu+1} \biggl(\partial_\nu P_\nu(x) \, P'_\nu(x)\biggr)_{x=a} .
\end{equation}
Some approximate formulas for computing $\partial_\nu P_\nu(a)$ were
also provided in that reference.

Hobson derived the following representation of $P_\nu(x)$ for
negative $x$ \cite{Hobson}:
\begin{widetext}
\begin{equation}  \label{eq:Hobson}
P_\nu(x) = \frac{\sin(\pi \nu)}{\pi} \biggl\{\biggl[\ln \biggl(\frac{1+x}{2}\biggr) + 2\gamma + \psi(\nu+1) + \psi(-\nu)\biggr] 
P_\nu(-x)
 + \sum\limits_{k=1}^\infty \frac{(-1)^k \Gamma(\nu+k+1)}{\Gamma(\nu-k+1) (k!)^2} \, \varphi(\nu,k) \biggl(\frac{1+x}{2}\biggr)^k \biggr\} ,
\end{equation}
\end{widetext}
where $\psi(\nu) = \frac{d \log \Gamma(\nu)}{d\nu}$ is the digamma
function, $\gamma = 0.5772\ldots$ is the Euler constant, and
\begin{eqnarray} \nonumber
\varphi(\nu,k) &=&  - 2 \biggl(\frac{1}{1} + \ldots + \frac{1}{k}\biggr) +
\biggl(\frac{1}{\nu+1} + \ldots + \frac{1}{\nu+k}\biggr)  \\  
&+& \biggl(\frac{1}{-\nu} + \ldots + \frac{1}{-\nu+k-1}\biggr) . 
\end{eqnarray}
Schelkunoff gave approximate expressions for Legendre functions of
nearly integral degree \cite{Schelkunoff41}, e.g.,
\begin{subequations}  \label{eq:Pnu_eps}
\begin{eqnarray}   
P_\epsilon(x) &\simeq& 1 + \epsilon \ln \biggl(\frac{1+x}{2}\biggr) , \\
P_{1+\epsilon}(x) & \simeq& x + \epsilon \biggl[x-1 + x \ln\biggl(\frac{1+x}{2}\biggr) \biggr],
\end{eqnarray}
\end{subequations}
which are applicable for all $|x|<1$ and yield an error of the order
of $\epsilon^2$.

To compute Legendre functions numerically at large $\nu$, it is
convenient to use the Mehler-Dirichlet integral representation
\begin{equation}  \label{eq:Pnu_int}
P_\nu(\cos\theta) = \frac{\sqrt{2}}{\pi} \int\limits_0^\theta d\alpha \frac{ \cos ((\nu+1/2)\alpha)}{\sqrt{\cos \alpha - \cos\theta}}  \,,
\end{equation}
which is valid for $\theta \in (0,\pi)$ and $\nu \in \C$
\cite{Hobson}.
From this representation, one can also evaluate
\begin{equation}  \label{eq:dPnu_int}
\partial_\nu P_\nu(\cos\theta) = - \frac{\sqrt{2}}{\pi} \int\limits_0^\theta d\alpha 
\frac{\alpha \sin ((\nu+1/2)\alpha)}{\sqrt{\cos \alpha - \cos\theta}} \,.
\end{equation}
These two relations were used throughout this paper for numerical
computations of $P_\nu(x)$ and $\partial_\nu P_\nu(x)$.

Note that the derivative of the Legendre function with respect to its
degree can also be written as \cite{DLMF,Szmytkowski06,Cohl11}:
\begin{equation}  \label{eq:dPnu_series}
\partial_\nu P_\nu(x) = \pi \ctan(\pi\nu) P_\nu(x) - \frac{1}{\pi} A_\nu(x) ,
\end{equation}
where
\begin{eqnarray}  \nonumber
A_\nu(x) &=& \sin(\pi\nu) \sum\limits_{k=0}^\infty \frac{((1-x)/2)^k \Gamma(k-\nu) \Gamma(k+\nu+1)}{(k!)^2} \\
&\times& \bigl(\psi(k+\nu+1) - \psi(k-\nu)\bigr) . 
\end{eqnarray}
One also gets
\begin{equation}  \label{eq:ddPnu_series}
\partial_\nu P'_\nu(a) = \frac{P'_\nu(a)}{\nu+1} + \frac{\nu+1}{a^2-1} \bigl[\partial_\nu P_{\nu+1}(a) - a \partial_\nu P_\nu(a)\bigr] .
\end{equation}
However, these relations are less convenient than the integral
presentation (\ref{eq:dPnu_int}) for numerical computations.

\section{Spectral properties}

\subsection{Solutions $\nu_n^s$ of Eq. (\ref{eq:Robin_P})}
\label{sec:spectral_lam0}

If $\nu$ and $\nu'$ denote any two solutions of Eq. (\ref{eq:Robin_P})
(with a fixed $q_s$), the right-hand side of
Eq. (\ref{eq:orthogonality}) is zero, ensuring the orthogonality of
the corresponding Legendre functions.  This property holds for Neumann
($q_s = 0$), Robin ($0 < q_s < \infty$) and Dirichlet ($q_s = \infty$)
boundary conditions.

Moreover, MacDonald used Eq. (\ref{eq:orthogonality}) to prove that
all zeros of $P_\nu(a)$ are real \cite{MacDonald1899} (see also an
amendment to the proof in \cite{Hobson}).  His argument can be
directly applied to prove that all zeros of $P'_\nu(a) - q_s P_\nu(a)$
are also real for any fixed real $q_s$.  It is enough to rewrite the
right-hand side of Eq. (\ref{eq:orthogonality}) as
\begin{equation*}
\frac{1}{q_s} \biggl( \bigl[q_s P_{\nu'}(a) - P'_{\nu'}(a)\bigr] P'_{\nu}(a) - \bigl[q_s P_{\nu}(a) - P'_{\nu}(a)\bigr] P'_{\nu'}(a) \biggr).
\end{equation*}
If $\nu$ was a complex zero of $P'_\nu(a) - q_s P_\nu(a)$, then its
complex conjugate, $\nu'= \nu^*$, would also be a zero (given that
$P_{\nu^*}(a) = P_\nu^*(a)$), implying that the integral in
Eq. (\ref{eq:orthogonality}) would vanish.  This is not possible
because $P_\nu(x) \, P_{\nu'}(x) = |P_\nu(x)|^2$ for such a pair of
complex conjugate degrees $\nu$ and $\nu'$ (see \cite{Hobson} for
details).

As $P_\nu(x)$ is an entire analytic function of $\nu$ for any fixed $x
> - 1$ (see Appendix \ref{sec:Legendre}), whereas $\nu^s$ is a
solution of Eq. (\ref{eq:Robin_P}) involving $P_\nu(a)$ and
$P'_\nu(a)$, we expect that $P_{\nu^s}(x)$ as a function of $s$ has no
poles.  A rigorous proof of this claim is beyond the scope of this
paper.

\subsection{Solutions $\nu_n$ of Eq. (\ref{eq:Robin_P2})}
\label{sec:spectral_lam}

The above arguments are not directly applicable for the solutions
$\nu_n$ of Eq. (\ref{eq:Robin_P2}).  In fact, rewriting again the
right-hand side of Eq. (\ref{eq:orthogonality}) in a form that is
consistent with Eq. (\ref{eq:Robin_P2}), one gets
\begin{eqnarray*}
&& \frac{(\nu-\nu')(\nu+\nu'+1)}{1-a^2}\int\limits_a^1 dx \, P_\nu(x) \, P_{\nu'}(x) \\
&& = P'_\nu(a) \biggl( P_{\nu'}(a) - \frac{\llambda+\mmu - \nu'(\nu'+1)}{q(\mmu - \nu'(\nu'+1))} P'_{\nu'}(a) \biggr) \\
&& - P'_{\nu'}(a) \biggl( P_{\nu}(a) - \frac{\llambda+\mmu - \nu(\nu+1)}{q(\mmu - \nu(\nu+1))} P'_{\nu}(a) \biggr) \\
&& -  \frac{\llambda (\nu-\nu')(\nu+\nu'+1)}{q(\mmu - \nu(\nu+1))(\mmu - \nu'(\nu'+1))} P'_{\nu}(a) P'_{\nu'}(a)\,.
\end{eqnarray*}
If $\nu$ and $\nu'$ are any two solutions of Eq. (\ref{eq:Robin_P2}),
the first two terms on the right-hand side vanish, but the last term
does not.  As a consequence, the functions $P_\nu(x)$ and
$P_{\nu'}(x)$, as well as all $P_{\nu_n}(x)$, determining $c(x,t)$ and
$J(t)$ via Eqs. (\ref{eq:vt}, \ref{eq:Jt}), are not orthogonal to each
other.

At the same time, the above relation still allows us to prove that
all solutions of Eq. (\ref{eq:Robin_P2}) are real.  If there existed a
complex solution $\nu$, then its complex conjugate, $\nu' = \nu^*$,
would also be a solution so that the above relation would become
\begin{equation*}
\int\limits_a^1 dx \, |P_\nu(x)|^2 = -  \frac{\llambda (1-a^2) |P'_{\nu}(a)|^2}{q |\mmu - \nu(\nu+1)|^2} \,.
\end{equation*}
However, this relation cannot hold because the left-hand side is the
integral of a positive function, whereas the right-hand side is
negative.

\subsection{Computation of the residues}
\label{sec:residues}

The Laplace inversion formulas include the factor $(\partial_s
f_n)(s_n)$ which can be written as
\begin{equation}  \label{eq:residue_a1}
\partial_s f_n(s_n) = 1 + \frac{D}{R^2} (2\nu_n + 1) (\partial_s \nu_n)(s_n) ,
\end{equation}
where $\nu_n$ is a shortcut notation for $\nu_n^{s_n}$, i.e., the
$n$-th solution of Eq. (\ref{eq:Robin_P}) evaluated at the pole $s_n$.
The last factor can be found by differentiating the boundary condition
(\ref{eq:Robin_P}):
\begin{equation}  \label{eq:residue_a2}
\partial_s \nu_n(s_n) = \frac{\kkappa \lambda \, P_{\nu_n}(a)
 \biggl( \partial_\nu P'_\nu(a) - q_{s_n} \partial_\nu P_\nu(a)\biggr)^{-1}_{\nu = \nu_n}}{\sqrt{1-a^2} (s_n+\lambda+\mu)^2}  \,,
\end{equation}
so that one needs to evaluate the derivative with respect to the
degree $\nu$.  Using Eqs. (\ref{eq:dPnu_series},
\ref{eq:ddPnu_series}) and the boundary condition (\ref{eq:Robin_P}),
one gets
\begin{align}    \label{eq:residue_a3}
& \biggl(\partial_\nu P'_\nu(a) - q_{s_n} \partial_\nu P_\nu(a)\biggr)_{\nu=\nu_n}  = \frac{P'_{\nu_n}(a)}{\nu_n+1}  \\  \nonumber
& - \frac{1}{\pi} \biggl[\frac{\nu_n+1}{a^2-1} \bigl(A_{\nu_n+1}(a) - a A_{\nu_n}(a)\bigr) - q_{s_n} A_{\nu_n}(a)\biggr] . 
\end{align}
In practice, we used the integral representation (\ref{eq:dPnu_int})
for evaluating Eq. (\ref{eq:residue_a2}).

\subsection{Estimates for zeros $\nu_n^s$}
\label{sec:Aestimates}

We discuss some properties of the solutions $\nu_n^s$ of
Eq. (\ref{eq:Robin_P}) (with fixed $a$ and $q_s$) that may facilitate
their numerical computation.  Even though most of the following
statements are expected to be classical, we could not find their
mathematical proofs in the literature.  As such proofs are beyond the
scope of the paper, most of the results of this section of the
Appendix remain conjectural from the mathematical point of view, even
so they have been checked numerically.  We emphasize that the
conjectural results of this section were not directly used in the main
text, except for guiding the author in the analysis.  As a
consequence, their conjectural character does not impair the quality
of main results.  In the following, we assume that $q_s \geq 0$.

Since the operator $\L = \partial_x (1-x^2) \partial_x$ is
self-adjoint, the minimax principle should imply the standard
inequalities between Dirichlet, Robin, and Neumann eigenvalues (see
\cite{Courant,Grebenkov13} for the case of the Laplace operator in
Euclidean domains), from which
\begin{equation}  \label{eq:inequal}
\nu_n^N \leq \nu_n^s \leq \nu_n^D ,
\end{equation}
where $\nu_n^D$ and $\nu_n^N$ are the zeros corresponding to the
limiting cases of Dirichlet ($q_s = \infty$) and Neumann ($q_s = 0$)
conditions:
\begin{equation}
P_{\nu_n^D}(a) = 0, \qquad  P'_{\nu_n^N}(a) = 0 .
\end{equation}
We also expect that $\nu_n^s$ are monotonous functions of $a$:
\begin{equation}
n = \nu_n^s(-1) \leq \nu_n^s(a_1) \leq \nu_n^s(a_2)  \qquad (a_1 < a_2) .
\end{equation}

Using the asymptotic expansions of the Legendre function at large
$\nu$ \cite{Abramowitz}
\begin{equation}
\begin{split}
P_\nu(\cos\theta) & = \frac{\Gamma(\nu+1)}{\Gamma(\nu+3/2)} \biggl(\frac{\pi \sin\theta}{2}\biggr)^{-1/2} \\
& \times \cos\bigl((\nu+1/2)\theta - \pi/4\bigr) + O(\nu^{-1}), \\
\end{split}
\end{equation}
we get
\begin{equation}
\nu_n^D \simeq \nu_n^{D,\rm app} = \frac{n + 3/4}{1 - \ve/\pi} - \frac12   \qquad (n\to\infty),
\end{equation}
where we used $\arccos(a) = \pi - \ve$.  Note that this asymptotic
relation is exact for $a = 0$ (or $\ve = \pi/2$): $\nu_n^D = 2n+1$.
We conjecture that for all $n = 0,1,2,\ldots$,
\begin{subequations}
\begin{eqnarray}
&& \nu_n^D \leq \nu_n^{D,\rm app}    \qquad (a < 0) ,\\
&& \nu_n^D \geq \nu_n^{D,\rm app}    \qquad (a > 0) .
\end{eqnarray}
\end{subequations}
In other words, the asymptotic approximation $\nu_n^{D,\rm app}$ is
either the upper bound (for $a<0$), or the lower bound (for $a>0$) for
the zeros $\nu_n^D$.  These inequalities have been checked numerically
for several values of $a$ and $n$.

Similarly, taking the derivative of $P_\nu(x)$ with respect to $x$,
one gets
\begin{eqnarray}
P'_\nu(\cos\theta) &=& \frac{\pi}{2} \frac{\Gamma(\nu+2)}{\Gamma(\nu+3/2)} \biggl(\frac{\pi \sin\theta}{2}\biggr)^{-3/2} \\  \nonumber
&\times& \sin\bigl((\nu+1/2)\theta - \pi/4\bigr) + O(\nu^{-1}), 
\end{eqnarray}
so that
\begin{equation}
\nu_n^N \simeq \nu_n^{N,\rm app} = \frac{n + 1/4}{1 - \ve/\pi} - \frac12   \qquad (n\to\infty).
\end{equation}
This asymptotic relation becomes exact for $a = 0$ (or $\ve = \pi/2$):
$\nu_n^N = 2n$.  We conjecture that for all $n = 0,1,2,\ldots$,
\begin{subequations}
\begin{eqnarray}
&& \nu_n^N \geq \nu_n^{N,\rm app}   \qquad (a < 0) ,\\
&& \nu_n^N \leq \nu_n^{N,\rm app}   \qquad (a > 0) .
\end{eqnarray}
\end{subequations}
In other words, the asymptotic approximation $\nu_n^{N,\rm app}$ is
either the lower bound (for $a<0$), or the upper bound (for $a>0$) for
the zeros $\nu_n^N$.  These inequalities have been checked
numerically for several values of $a$ and $n$.

Combining these inequalities with Eq. (\ref{eq:inequal}), we get the
following conjectural inequalities for $a < 0$:
\begin{equation}
\nu_n^{N,\rm app} \leq \nu_n^{N} \leq \nu_n^s \leq \nu_n^D \leq \nu_n^{D,\rm app} .
\end{equation}
We note that the case $a < 0$ (small targets) is more relevant for
applications than $a > 0$ (large targets).  These inequalities allow
one to impose constraints on the range of $\nu$, on which the $n$-th
zero $\nu_n^s$ is searched numerically.  At $n = 0$, the lower bound
is negative and should thus be replaced by $0$.

\section{Example of numerical implementation}
\label{sec:Anumerics}

In this Appendix, we illustrate the computation of the total flux
$J(t)$ shown in Fig. \ref{fig:nu0}(b).  After truncation of the
spectral decomposition (\ref{eq:Jt}) at $\nmax = 20$, this flux is
approximated by 20 exponential functions, determined by $\nu_n$ and
$J_n$, with $n = 0,1,2,\ldots,19$.  For a given set of dimensionless
parameters ($\ve$, $\bar{\kappa}$, $\bar{\lambda}$, and $\bar{\mu}$),
we first determined $\nu_n$ by solving Eq. (\ref{eq:Robin_P2}) by a
bisection method; then the coefficients $J_n$ were computed from
Eq. (\ref{eq:Jn}), as discussed in Sec. \ref{sec:numerics}.

\begin{figure}
\begin{center}
\includegraphics[width=88mm]{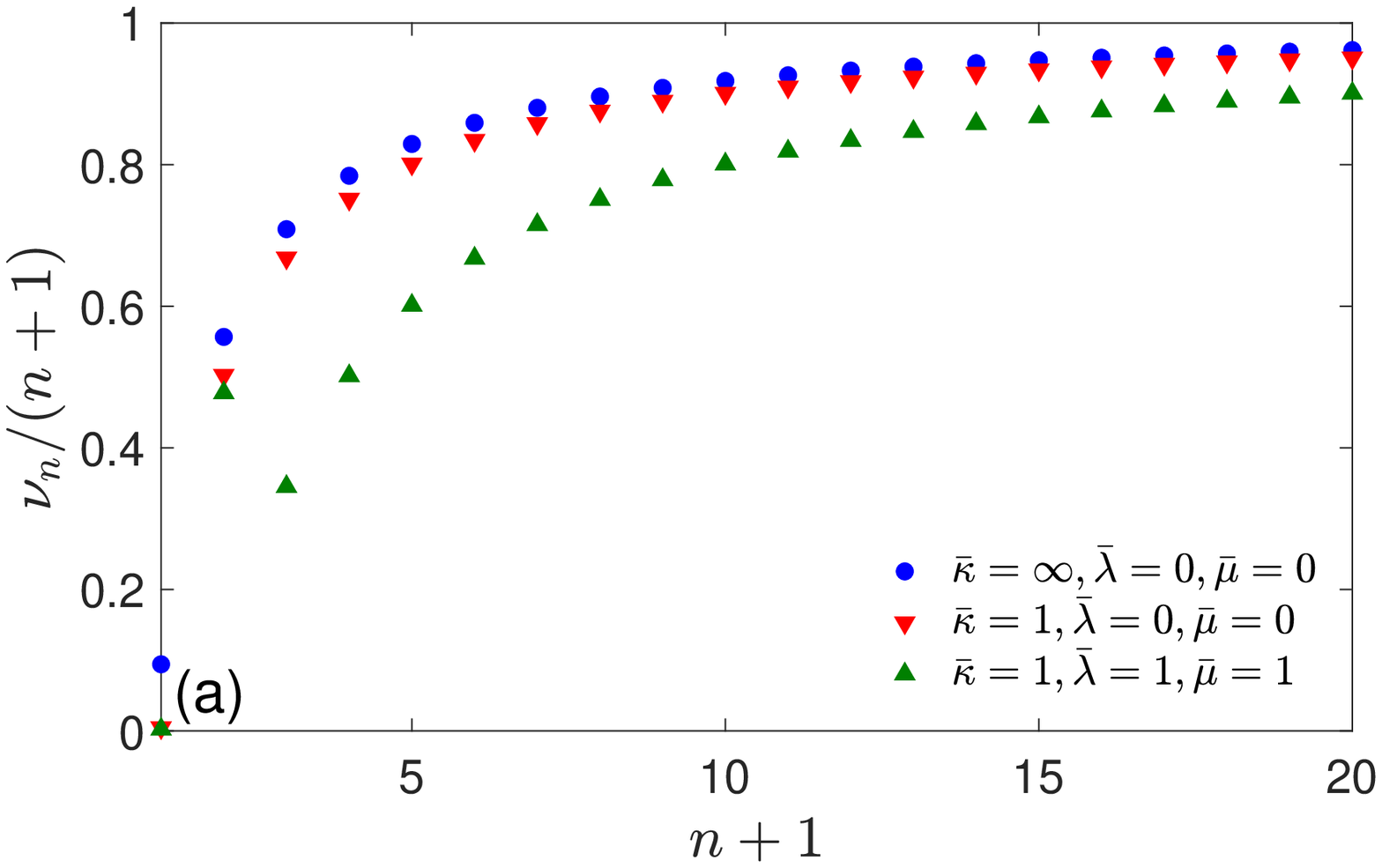} 
\includegraphics[width=88mm]{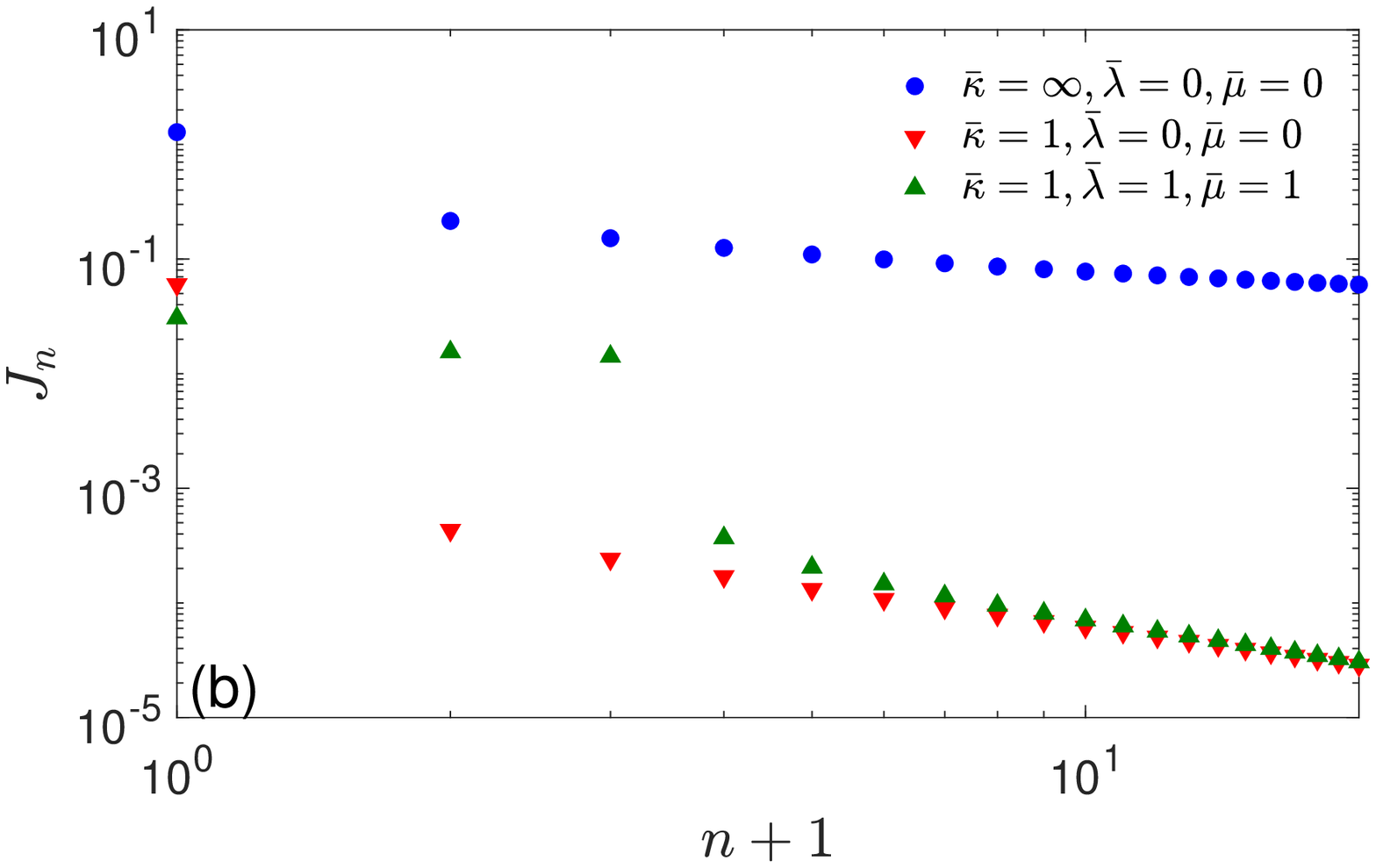} 
\end{center}
\caption{
Parameters used to plot Fig. \ref{fig:nu0}(b) with $\ve = 0.01$ for
three configurations of reaction parameters: perfectly reactive target
($\bar{\kappa} = \infty$), partially reactive target with irreversible
binding ($\bar{\kappa} = 1$, $\bar{\lambda} = 0$), and partially
reactive target with partly reversible binding ($\bar{\kappa} = 1$,
$\bar{\lambda} = 1$, $\bar{\mu} = 1$).  {\bf (a)} The first 20
solutions $\nu_n$ of Eq. (\ref{eq:Robin_P2}), divided by $n+1$ for
convenience.  {\bf (b)} The first 20 coefficients $J_n$ in
Eq. (\ref{eq:Jt}) for the total flux $J(t)$. }
\label{fig:nun}
\end{figure}

Figure \ref{fig:nun} shows the first 20 parameters $\nu_n$ and $J_n$
that were used for plotting Fig. \ref{fig:nu0}(b) for a small target
$\ve = 0.01$ and three configurations of reaction parameters.  For
convenience, $\nu_n$ are divided by $n+1$, to highlight their
asymptotic behavior $\nu_n \propto n$.  In all three cases, the
smallest solution $\nu_0$ is small but strictly positive (its
dependence on $\ve$ was shown in Fig. \ref{fig:nu0}(a)).  As $n$
increases, $\nu_n/(n+1)$ approaches $1$ from below, suggesting that
$n+1$ is the upper bound of $\nu_n$.  Even though a rigorous proof of
this statement is beyond the scope of the paper (see also Appendix
\ref{sec:Aestimates}), one can use the relation $\nu_n \propto n$ to
estimate the validity range of the truncated decomposition.  In fact,
the omission of the terms with $n \geq \nmax$ is justified when $Dt
\nu_n(\nu_n+1)/R^2 \gg 1$, i.e., $Dt/R^2 \gg \nmax^{-2}$.  This
inequality is satisfied for all figures in the paper.

For the considered three cases of reaction parameters, the
coefficients $J_n$ are positive and decreasing with $n$.  The first
coefficient $J_0$ provides the dominant contribution to the total
flux, particularly for the case of a partial sink with irreversible
reaction ($\bar{\kappa} = 1$, $\bar{\lambda} = \bar{\mu} = 0$).

\section{Derivation of the MFPT}
\label{sec:MFPT_Robin}

The mean FPT to a partially reactive target on the surface of a sphere
was found in \cite{Sano81}.  Here we reproduce its computation for the
sake of completeness.  The MFPT $\tau(x)$ obeys the Poisson equation:
$D \Delta \tau = -1$ that reads as
\begin{equation}
\partial_x (1-x^2) \partial_x \tau(x) = - \frac{R^2}{D} \qquad (a < x < 1),
\end{equation}
subject to the Robin boundary condition at $x=a$: $\frac{D}{R}\sin\ve
(\partial_x \tau)(a) = \kappa \tau(a)$.  A general solution is
obtained by integrating twice the above equation:
\begin{equation}
\tau(x) = \frac{R^2}{2D} \ln(1-x^2) + \frac{c_1}{2} \ln \frac{1+x}{1-x} + c_2 ,
\end{equation}
with two constants $c_1$ and $c_2$.  To ensure the regularity of this
solution at $x=1$, we set $c_1 = R^2/D$ so that
\begin{equation}
\tau(x) = \frac{R^2}{D} \ln(1+x) + c_2 .
\end{equation}
The constant $c_2$ is fixed by the Robin boundary condition, from
which Eq. (\ref{eq:MFPT_Robin}) follows.

\section{Asymptotic behavior for small targets}
\label{sec:asympt}

In this Appendix, we investigate separately the asymptotic behavior of
the solutions $\nu_n^s$ of Eq. (\ref{eq:Robin_P}) and that of the
solutions $\nu_n$ of Eq. (\ref{eq:Robin_P2}) in the limit $\ve \to 0$.

\subsection{Solutions $\nu_n^s$ of Eq. (\ref{eq:Robin_P})}

Here we analyze the asymptotic behavior of the solutions $\nu_n^s$ of
Eq. (\ref{eq:Robin_P}) for a fixed $q_s$.  

For small $\ve$, we have $a \simeq -1 + \ve^2/2$.  Using the Hobson's
representation (\ref{eq:Hobson}) and $P_\nu(-a) = P_\nu(1-\ve^2/2) = 1
+ O(\ve^2)$, we get
\begin{equation}
P_\nu(a) \simeq \frac{\sin(\pi\nu)}{\pi}\, \Psi_\ve(\nu) + O(\ve^2) ,
\end{equation}
with
\begin{equation}  \label{eq:Psinu}
\Psi_\ve(\nu) = \ln(\ve^2/4) + 2\gamma + 2\psi(\nu+1) + \pi \ctan(\pi \nu),
\end{equation}
where we used the reflection property of the digamma function:
$\psi(1+z) - \psi(-z) = -\pi \ctan(\pi \nu)$.  Substituting this
expression into Eq. (\ref{eq:Robin_P1a}) (which is equivalent to
Eq. (\ref{eq:Robin_P})), we get
\begin{equation}  \label{eq:auxil33}
\pi \ctan(\pi \nu) + 2\psi(\nu+1) \simeq \frac{2}{q_s \ve^2} + 2\ln(2/\ve) - 2\gamma ,
\end{equation}
where we used that $\psi(z+1) = \psi(z) + 1/z$.  Since the right-hand
side of this relation is large as $\ve\to0$, $\nu$ should lie near the
poles of $\ctan(\pi \nu)$.  We search thus an approximate solution of
Eq. (\ref{eq:auxil33}) as
\begin{equation}
\nu_n^s \simeq n + \delta_n, \qquad \delta_n \ll 1.
\end{equation}
Note that $\psi(\nu+1)$ is regular at positive integers: $\psi(n+1) =
H_n - \gamma$, where $H_n = 1/1 + \ldots + 1/n$ are the harmonic
numbers (and $H_0 = 0$).  One gets thus
\begin{equation}  \label{eq:deltan}
\frac{1}{\delta_n} \simeq \frac{2}{q_s \ve^2} + 2\ln(2/\ve) - 2H_n .
\end{equation}
This approximate solution is valid only when the right-hand side is
large (and $\delta_n$ is small), yielding a constraint of having not
too large $n$.

In what follows, we will need to evaluate $P'_{\nu_n^s}(a)$.  Using
again Eq. (\ref{eq:Hobson}), we find in the leading term
\begin{equation}
P'_{\nu_n^s}(a) \simeq \frac{\sin (\pi \nu_n)}{\pi(1+a)} \simeq \frac{2\delta_n}{1-a^2} \,.
\end{equation}
Using Eqs. (\ref{eq:residue_a1}, \ref{eq:residue_a2},
\ref{eq:residue_a3}) and representations (\ref{eq:dPnu_series},
\ref{eq:ddPnu_series}), we also determine 
\begin{equation}
(\partial_s f_n)(s_n) \simeq 1  \qquad (\ve\to 0).
\end{equation}
Finally, given that $b_n^2 = n+1/2$ for $\ve = 0$, this relation
should hold approximately for small $\ve$.

In the following, we consider separately the cases of perfect ($\kappa
= \infty$) and imperfect ($\kappa < \infty$) target.

\subsubsection*{Perfect sink}

For a perfect sink ($\kappa = \infty$), one has $q_s = \infty$, and
the first term in Eq. (\ref{eq:deltan}) vanishes, so that one
retrieves the asymptotic formula for the zeros of the Legendre
function from Ref. \cite{Hall49}:
\begin{equation}
\nu_n^s = \nu_n \simeq n + \frac{1}{2\ln(2/\ve) - 2H_n} \,.
\end{equation}
For $n > 0$, the logarithmic term provides a small correction and can
thus be neglected so that $\nu_n \simeq n$.  In contrast, when $n =
0$, the logarithmic term dominates.  One gets thus the asymptotic
behavior of $\tilde{c}(x,s)$ and $\tilde{J}(s)$ via
Eqs. (\ref{eq:tildev}, \ref{eq:tildeJ}).

Moreover, as $q_s$ is infinite (and thus does not depend on $s$), the
degrees $\nu_n$ determine directly the poles via Eq. (\ref{eq:sn}).
In particular, one gets
\begin{equation}
s_0 \simeq - \frac{D}{2R^2\ln(2/\ve)} \,.
\end{equation}
Note that $1/|s_0|$ is close to the mean first passage time to the
target in this domain (see Appendix \ref{sec:MFPT_Robin}).
Substituting these results into Eq. (\ref{eq:Jt}), we get the
long-time behavior (\ref{eq:Jtinf}) of the total flux.

\subsubsection*{Imperfect sink}

For imperfect sink ($\kappa < \infty$), one can neglect the
logarithmic and constant terms in Eq. (\ref{eq:deltan}), so that 
\begin{equation}
\nu_n^s \simeq n + \frac{\kkappa (s + \mu)}{2(s + \lambda + \mu)} \,\ve + o(\ve) .
\end{equation}
This asymptotic relation determines $\tilde{c}(x,s)$ and
$\tilde{J}(s)$ via Eqs. (\ref{eq:tildev}, \ref{eq:tildeJ}).

Moreover, for irreversible reactions with $\lambda = 0$, $q_s$ and
thus $\nu_n^s$ do not depend on $s$ (and $\mu$) so that one can use
the above relation to determine the asymptotic behavior of $c(x,t)$
and $J(t)$ in Eqs. (\ref{eq:vt}, \ref{eq:Jt}).  In fact,
Eq. (\ref{eq:sn}) yields
\begin{equation}
s_n = - \frac{D}{R^2} \biggl(n(n+1) + (n+1/2) \ve \kkappa + O(\ve^2)\biggr) .
\end{equation}
In a first approximation, one can neglect $\ve$ in the higher-order
eigenmodes.  Keeping only the term with $n = 0$ with the smallest rate
$|s_0|$, we get the long-time behavior (\ref{eq:Jtinf}) of the total
flux.

\subsection{Solutions $\nu_n$ of Eq. (\ref{eq:Robin_P2})}

For reversible reactions (with $\lambda > 0$), $c(x,t)$ and $J(t)$ are
determined by the solutions $\nu_n$ of Eq. (\ref{eq:Robin_P2}).  To
get their asymptotic behavior in the small-target limit, we rewrite
Eq. (\ref{eq:Robin_P2}) as
\begin{equation} 
\llambda + \mmu - \nu(\nu+1) = \kkappa (\mmu - \nu(\nu+1)) \, \frac{P_\nu(a)}{\sqrt{1-a^2} \, P'_\nu(a)} \,.
\end{equation}
The Hobson representation (\ref{eq:Hobson}) for $P_\nu(a)$ and its
derivative $P'_\nu(a)$ yields in the limit $\ve \to 0$:
\begin{equation} \label{eq:auxil55}
\llambda + \mmu - \nu(\nu+1) =  \ve \, \frac{\kkappa (\mmu - \nu(\nu+1)) \bigl[\Psi_\ve(\nu) + O(\ve^2)\bigr]}
{2 + \ve^2 \frac{\nu(\nu+1)}{2} \Psi_\ve(\nu) + O(\ve^2)} \,,
\end{equation}
with $\Psi_\ve(\nu)$ given by Eq. (\ref{eq:Psinu}).  As the function
$\Psi(\nu)$ diverges at any integer $\nu$, one can search again for
solutions near integers, $\nu_n = n + \delta_n$, from which 
\begin{equation}
\Psi_\ve(\nu_n) = \frac{1}{\delta_n} + \ln(\ve^2/4) + 2H_n + O(\delta)
\end{equation}
and thus
\begin{equation}  \label{eq:deltan_eps}
\delta_n \simeq  \frac{\kkappa(\mmu - n(n+1))}{2(\llambda+\mmu -n(n+1))} \, \ve + o(\ve).
\end{equation}
For instance, one gets the following asymptotic relation for the
smallest degree:
\begin{equation}
\nu_0 = \delta_0 \approx \frac{\kkappa \, \mmu \, \ve}{2(\mmu + \llambda)} \approx \frac{\mu \kappa \rho}{2D(\mu + \lambda)} \,.
\end{equation}
In the particular case when $\llambda+\mmu = n'(n'+1)$ for some
integer $n'$, Eq. (\ref{eq:deltan_eps}) does not hold for this $n'$,
and one gets a slower approach to the limit:
\begin{equation}
\delta_{n'} \simeq \biggl(\frac{\kkappa \, \llambda}{2(2n'+1)}\biggr)^{1/2}  \, \sqrt{\ve} + o(\sqrt{\ve}).
\end{equation}

Yet another solution of Eq. (\ref{eq:auxil55}) is possible when its
left-hand side is of the order of $\ve$ and the corresponding $\nu$ is
not close to an integer.  In this case, $\Psi_\ve(\nu)$ behaves
asymptotically as $2\ln(\ve/2) + O(1)$ and thus one gets the solution
in the leading order in $\ve$:
\begin{equation}  \label{eq:nu_special}
\nu(\nu+1) = \llambda + \mmu - \kkappa \llambda \, \ve \ln (1/\ve) + O(\ve) .
\end{equation}
As a consequence, the degree $\nu$ approaches the value $\sqrt{1/4 +
\llambda+\mmu}-1/2$ in the limit $\ve \to 0$.  We recall that this
value should not be integer (otherwise the relation
(\ref{eq:nu_special}) does not necessarily hold).  This solution
yields the eigenvalue, which is close to $\llambda + \mmu$.  We
emphasize that the above argument is not applicable when $\lambda =
0$, in which case $q_s$ does not depend on $s$ any more.

\end{document}